\documentclass[11pt,a4paper]{article}

\usepackage{jheppub}

\usepackage[scr=boondox]{mathalpha}
\usepackage{array}
\usepackage{mathtools}
\usepackage{wasysym}
\usepackage{slashed}
\usepackage{bbm}
\usepackage{bm}
\usepackage{multirow}
\usepackage{booktabs}
\usepackage{orcidlink}
\usepackage[normalem]{ulem}

\usepackage[caption=false]{subfig}
\usepackage{svg}

\usepackage{xcolor}
\definecolor{Nonoha}{HTML}{7DC7B0}
\definecolor{Shino}{HTML}{BD81D4}
\definecolor{Mikaru}{HTML}{FFE673}
\definecolor{Miku}{HTML}{FF8C73}
\definecolor{Chima}{HTML}{80C8EF}
\definecolor{Toko}{HTML}{92F3A4}
\usepackage{tikz}
\usepackage[compat=1.1.0]{tikz-feynman}
\usepackage[compat=1.1.0]{tikz-feynhand}
\usetikzlibrary{plotmarks, shapes}

\newcommand{\sumint}{\mathop{\mathchoice
  {\mathrlap{\mkern8mu\int}\sum}%
  {\mathrlap{\mkern5mu\vcenter{\hbox{\scalebox{1.25}{$\int$}}}}\sum}%
  {\mathrlap{\mkern8mu\int}\sum}%
  {\mathrlap{\mkern8mu\int}\sum}%
}\nolimits}

\newcommand{\ii}{\mathrm{i}}

\allowdisplaybreaks

\title{Three-body unitary determination of the {\boldmath$f_1(1285)$} and {\boldmath$f_1(1420)$} pole positions}

\author[a,b]{Tao-Ran Hu\,\orcidlink{0009-0003-9720-0171},}
\author[c,d]{Hai-Long Fu\,\orcidlink{0000-0002-1722-4145},}
\author[b,a,e]{Feng-Kun Guo\,\orcidlink{0000-0002-2919-2064},}
\author[c,d,f]{Ulf-G.~Mei{\ss}ner\,\orcidlink{0000-0003-1254-442X},}
\author[b]{and Xu Zhang\,\orcidlink{0000-0002-3687-248X}}

\affiliation[a]{
    School of Physical Sciences, University of Chinese Academy of Sciences, Beijing 100049, China}
\affiliation[b]{
    Institute of Theoretical Physics, Chinese Academy of Sciences, Beijing 100190, China}
\affiliation[c]{
    Helmholtz-Institut für Strahlen- und Kernphysik, Bethe Center for Theoretical Physics and\\ Cluster of Excellence ``Color meets Flavor'', Universität Bonn, D-53115 Bonn, Germany}
\affiliation[d]{
    Institute for Advanced Simulation (IAS-4), Forschungszentrum Jülich, D-52425 Jülich, Germany}
\affiliation[e]{
    Southern Center for Nuclear-Science Theory (SCNT), Institute of Modern Physics, Chinese Academy of Sciences, Huizhou 516000, China}
\affiliation[f]{
    Peng Huanwu Collaborative Center for Research and Education, Beihang University, Beijing 100191, China}

\emailAdd{hutaoran21@mails.ucas.ac.cn, h.fu@fz-juelich.de, fkguo@itp.ac.cn, meissner@hiskp.uni-bonn.de, zhangxu@itp.ac.cn}

\abstract{
We study the $I^G(J^{PC})=0^+(1^{++})$ $K\bar K\pi$ system in an infinite-volume three-body unitary framework, focusing on the pole content of the region of the $f_1(1285)$ and $f_1(1420)$ resonances. The coupled $\pi a_0$-$K\bar K^*$ amplitude is constructed in the spectator-isobar representation, where the one-particle-exchange interaction required by three-body unitarity automatically incorporates the triangle-singularity mechanism. The short-range three-body interaction is constrained by fitting the $0^+(1^{++})$ component of the BESIII $K^0_SK^0_S\pi^0$ invariant-mass distribution in the $J/\psi\to\gamma(K^0_SK^0_S\pi^0)$ decay. Analytically continuing the fitted amplitude to the relevant unphysical Riemann sheets, we find two robust poles:
\begin{align}
\sqrt{s_{f_1(1285)}}&=
\left(1277\pm2\pm1\right)
-\ii\left(12\pm1\pm0\right)\text{MeV}\,,\notag\\
\sqrt{s_{f_1(1420)}}&=
\left(1435\pm2\pm7\right)
-\ii\left(40\pm2\pm1\right)\text{MeV}\,.\notag
\end{align}
The pole trajectories indicate that the $f_1(1285)$ originates from dressing a bare state introduced in the potential. In contrast, the $f_1(1420)$ is predominantly dynamically generated, and a single-channel analysis traces it to an $S$-wave $K\bar K^*$ quasi-bound state mixed with the nearby bare state, supporting its hadronic-molecule interpretation. We also find an additional pole deeper in the complex plane in the best-fit amplitude on the same Riemann sheet as the $f_1(1285)$. This additional pole is generated by the $P$-wave $\pi a_0$ contact interaction alone. It has a sizable cutoff and two-body-input dependence, and leaves little visible imprint on the physical lineshape. Finally, we provide a detailed and pedagogical appendix on how three-body cuts affect the solution of the integral equation.
}

\begin{document}

\maketitle

\section{Introduction}

Three-body dynamics is a long-standing problem in hadron physics. Many hadron resonances are observed through decays into three or more particles, and in several important cases the nearby two-body subsystems are themselves resonant. The separation between a genuine quantum chromodynamics (QCD) state and a kinematic enhancement is then not always transparent from a lineshape alone. Examples range from light-meson systems~\cite{COMPASS:2009xrl, Wu:2011yx, Aceti:2012dj, Davier:2013sfa, COMPASS:2015kdx, Mikhasenko:2015oxp, Aceti:2016yeb, COMPASS:2018uzl, Du:2019idk, COMPASS:2020yhb, COMPASS:2021ogp} and baryon excitations with large $\pi\pi N$ components~\cite{Arndt:1995bj, Kamano:2008gr, MartinezTorres:2008kh, Doring:2009yv, Alvarez-Ruso:2010ayw, Ronchen:2012eg, Kamano:2026vse} to near-threshold charmonium-like structures~\cite{BESIII:2013ris, Belle:2013yex, Wang:2013cya, Albaladejo:2015lob, Pilloni:2016obd, Gong:2016jzb}; see also Refs.~\cite{Guo:2019twa, JPAC:2021rxu, Liu:2024uxn, Doring:2025sgb} for recent reviews. Even for genuine resonances, such as the $X(3872)$~\cite{Belle:2003nnu, Baru:2011rs, LHCb:2013kgk} and $T_{cc}^+$~\cite{LHCb:2021vvq, LHCb:2021auc, Du:2021zzh, Lin:2022wmj, Zhang:2024dth, Hansen:2024ffk, Dawid:2024dgy, Dawid:2025wsn} in the hidden-charm and doubly charmed sectors, their couplings to $D\bar D\pi$ or $DD\pi$ channels are crucial to determine the physical lineshapes and the pole positions.

Several complementary approaches have been developed to treat such three-body effects. Dispersive methods and Khuri-Treiman equations have been extensively used to describe three-meson final-state interactions in light-hadron decays~\cite{Khuri:1960zz, Kambor:1995yc, Schneider:2010hs, Kampf:2011wr, Niecknig:2012sj, Danilkin:2014cra, Guo:2015zqa}. Non-relativistic Lippmann-Schwinger equations are particularly useful for heavy near-threshold systems, where the relevant momenta are small and unstable two-body constituents can be included explicitly~\cite{Baru:2011rs, Zhang:2020mpi, Du:2021zzh, Ji:2022blw, Zhang:2023wdz, Zhang:2024fxy, Ji:2025hjw, Fu:2025joa}. Relativistic field-theoretic formulations and finite-volume three-body quantization conditions provide another route, especially for connecting lattice-QCD spectra to infinite-volume amplitudes~\cite{Hansen:2014eka, Hansen:2015zga, Briceno:2017tce, Jackura:2020bsk, Dawid:2023jrj, Hansen:2024ffk, Dawid:2024dgy}. The approach adopted in this work is the infinite-volume unitary amplitude in the spectator-isobar representation, in which the amplitude is constrained by three-body unitarity and can be analytically continued to search for resonance poles~\cite{Mai:2017vot, Sadasivan:2020syi, Sadasivan:2021emk, Feng:2024wyg, Sakthivasan:2024uwd, Doring:2025phq, Sakthivasan:2026bph}; the corresponding finite-volume unitarity method can also be utilized in lattice-QCD studies~\cite{Mai:2017bge, Doring:2018xxx, Mai:2018djl, Mai:2021nul, Yan:2024gwp, Yan:2025mdm, Feng:2026ixm}; see Ref.~\cite{Meissner:2022cbi} for a textbook overview. The relation between different formulations of three-body dynamics has been discussed in Refs.~\cite{Aitchison:1966lpz, Aitchison:2015jxa, Jackura:2018xnx, Hansen:2019nir, Jackura:2019bmu, Blanton:2020jnm, Mai:2021lwb, Meissner:2022cbi, Dawid:2025wsn}.

In this paper we apply the infinite-volume unitary framework to the $I^G(J^{PC})=0^+(1^{++})$ $K\bar K\pi$ system, with particular emphasis on the two structures historically known as the $f_1(1285)$ and $f_1(1420)$. Both states have played an important role in the spectroscopy of axial-vector mesons. Experimentally, the $f_1(1285)$ is a well-established narrow state~\cite{ParticleDataGroup:2026mpi}, while the $f_1(1420)$ appears close to the $K\bar K^*$ threshold and is observed dominantly in the $K\bar K\pi$ channel~\cite{Athens-Bari-Birmingham-CERN:1984kwn, WA76:1989zfi, WA76:1992tut, WA102:1997gkz, L3:2000gjc, E852:2001ote, OBELIX:2002eai, DELPHI:2003bnm, L3:2007obw}. In the $\pi a_0\to\eta\pi\pi$ channel, the experimental situation is more subtle: a clear $f_1(1285)$ signal is accompanied by additional strength around $1.4$~GeV, which has sometimes been associated with the $f_1(1420)$~\cite{WA102:1998zhh}.

The theoretical interpretation of these structures is not settled. In conventional quark models, the $f_1(1285)$ and $f_1(1420)$ are assigned to the same $1^{++}$ nonet~\cite{Godfrey:1985xj, Close:1997nm, Chen:2015iqa}. On the other hand, because the $f_1(1420)$ lies very close to the $K\bar K^*$ threshold and couples strongly to that channel, it has also been interpreted as a hadronic molecule~\cite{Longacre:1990uc, Tornqvist:1993ng}. A different conclusion was reached in unitarized chiral approaches, where the $f_1(1285)$, rather than the $f_1(1420)$, can emerge as a dynamically generated $K\bar K^*$ state~\cite{Lutz:2003fm, Roca:2005nm, Zhou:2014ila}.\footnote{More recently, it has also been argued that a virtual-state-like pole associated with the $K\bar K^*$ interaction may appear near threshold and contribute to the $f_1(1420)$ enhancement within a similar chiral unitary framework~\cite{Yan:2025bez}.} In that picture, the enhancement near $1.4$~GeV in the $K\bar K^*$ channel may arise from the $f_1(1285)$ tail together with the opening of the $K\bar K^*$ channel~\cite{Debastiani:2016xgg} (which is, however, inconsistent with the strongly enhanced sharp peaks observed), while that in the $\pi a_0$ channel is attributed to the triangle singularity mechanism.

A triangle singularity is a logarithmic branch point that occurs when the three internal particles in a triangle diagram can simultaneously go on shell and move collinearly so that the process can be interpreted as a classical space-time process~\cite{Landau:1959fi, Coleman:1965xm, Bayar:2016ftu}. It can produce a resonance-like peak when it is close to the physical region; see Ref.~\cite{Guo:2019twa} for a detailed review and, e.g., Refs.~\cite{Achasov:1987ku,Anisovich:1995zu,Wu:2011yx} for some early applications. For the present quantum numbers, the relevant mechanism is $f_1(1285)\to K\bar K^*$, followed by $\bar K^*\to \bar K\pi$ and $K\bar K\to a_0$, producing a $K\bar K \bar K^*$ triangle loop contribution to the $\pi a_0$ final state. Such a singularity is expected to generate a non-trivial structure in the lineshape near the $K\bar K^*$ threshold~\cite{Wu:2012pg, Aceti:2015zva, Aceti:2015pma, Achasov:2016wll, Debastiani:2016xgg, Xie:2019iwz, Du:2021zdg, Du:2022nno}. Therefore, a meaningful interpretation of the region of the $f_1(1285)$ and $f_1(1420)$ should not simply add or remove Breit-Wigner resonances, but should ask whether physical poles remain after the kinematic effects required by three-body dynamics are included.\footnote{A similar strategy has been used in studies of the $Z_c(3900)$ region, where one tests whether a pole persists after accounting for the relevant threshold and triangle-singularity effects~\cite{Albaladejo:2015lob,Pilloni:2016obd,Du:2022jjv, Chen:2026fnz}.}

The infinite-volume unitary framework used here is well suited for this purpose, because the one-particle-exchange kernel required by three-body unitarity automatically generates the corresponding triangle singularities~\cite{Sakthivasan:2024uwd}. We construct the coupled $\pi a_0$-$K\bar K^*$ amplitude in the $0^+(1^{++})$ sector, fit the short-range three-body interaction to the BESIII $K^0_SK^0_S\pi^0$ invariant-mass distribution in $J/\psi\to\gamma(K^0_SK^0_S\pi^0)$~\cite{BESIII:2022chl},\footnote{Refs.~\cite{Nakamura:2022rdd, Nakamura:2023hbt} also analyzed the same BESIII data employing a three-body unitary approach for the $\eta(1405/1475)$ system.} and analytically continue the resulting amplitude to the relevant unphysical Riemann sheets (RSs). This allows us to determine whether the structures associated with the $f_1(1285)$ and $f_1(1420)$ correspond to poles and to study their coupling pattern in the $\pi a_0$ and $K\bar K^*$ channels as well as their origins.

The paper is organized as follows. In Sec.~\ref{sec:formalism}, we present the infinite-volume three-body unitary formalism, including the spectator-isobar decomposition, the channel basis, the one-particle-exchange interaction, and the partial-wave projection. In Sec.~\ref{sec:inputs}, we specify the two-body and three-body physical inputs and construct the decay amplitude used for the lineshape fit. Section~\ref{sec:res} contains the fit results, the analytic continuation, the pole determination, the channel couplings, and the analysis of the origins of the poles. A technical discussion of the contour deformation method for real-axis decay amplitudes is given in Appendix~\ref{app:CD}.

\section{Three-body unitary amplitude}\label{sec:formalism}

\subsection{Bethe-Salpeter equation}

\begin{figure}[tb]
    \centering
    \[
\scalebox{0.6}
{
\begin{tikzpicture}[baseline=0cm, line cap=round, line join=round]
  \begin{feynman}[baseline=(a0d)]
  \begin{feynhand}
    \vertex []                         (a0){};
    \vertex [below=1cm of a0]          (ad){};
    \vertex [above=1cm of a0]          (au){};

    \vertex [right=2cm of a0]          (b0){};
    \vertex [below=1cm of b0]          (bd){};
    \vertex [above=1cm of b0]          (bu){};

    \vertex [right=2cm of b0]          (x0){};
    \vertex [below=1cm of x0]          (xd){};
    \vertex [above=1cm of x0]          (xu){};
    \propag[thick]              (au) to (xu);
    \propag[thick]              (a0) to (x0);
    \propag[thick]              (ad) to (xd);
    \filldraw[draw=Nonoha!75!black, fill=Nonoha!10, line width=0.45mm, rounded corners=2pt]
      (1.15,-1.15) rectangle (2.85,1.15)
      node[pos=0.5]  {\LARGE $\mathcal T_3$};
 \end{feynhand}
 \end{feynman}
\end{tikzpicture}
}
=
\scalebox{0.6}
{
\begin{tikzpicture}[baseline=0cm, line cap=round, line join=round]
  \begin{feynman}[baseline=(ad)]
  \begin{feynhand}
    \vertex []                         (a0){};
    \vertex [below=1cm of a0]          (ad){};
    \vertex [above=1cm of a0]          (au){};

    \vertex [right=0.7cm of a0]        (b0){};
    \vertex [below=1cm of b0]          (bd){};
    \vertex [above=1cm of b0,dot]      (bu){};

    \vertex [right=0.7cm of b0]        (x0){};
    \vertex [below=1cm of x0]          (xd){};
    \vertex [above=1cm of x0]          (xu){};
    \propag[double, thick] (bu) to (xu);
    \propag[thick]         (ad) to (xd);
    \propag[thick]         (au) to (bu);
    \propag[thick]         (a0) to (bu);
  \end{feynhand}
  \end{feynman}
\end{tikzpicture}
}
\left(
\scalebox{0.6}
{
\begin{tikzpicture}[baseline=0cm, line cap=round, line join=round]
  \begin{feynman}[baseline=(a0)]
  \begin{feynhand}
    \vertex []                         (a0){};
    \vertex [below=1cm of a0]          (ad){};
    \vertex [above=1cm of a0]          (au){};

    \vertex [right=0.7cm of a0]        (b0){};
    \vertex [below=1cm of b0]          (bd){};
    \vertex [above=1cm of b0,dot]      (bu){};

    \vertex [right=0.7cm of b0]        (x0){};
    \vertex [below=1cm of x0]          (xd){};
    \vertex [above=1cm of x0]          (xu){};
    \propag[double, thick] (au) to (xu);
    \propag[thick]         (ad) to (xd);
    \node[draw=Shino!75!black, fill=Shino!10, line width=0.32mm,
          rounded corners=1.5pt, minimum width=0.68cm, minimum height=0.54cm,
          inner sep=1.5pt] at (0.7,1) {\Large$\tau$};
  \end{feynhand}
  \end{feynman}
\end{tikzpicture}
}
+
\scalebox{0.6}
{
\begin{tikzpicture}[baseline=0cm, line cap=round, line join=round]
  \begin{feynman}[baseline=(a0d)]
  \begin{feynhand}
    \vertex []                         (a0){};
    \vertex [below=1cm of a0]          (ad){};
    \vertex [above=1cm of a0]          (au){};

    \vertex [right=2cm of a0]          (b0){};
    \vertex [below=1cm of b0]          (bd){};
    \vertex [above=1cm of b0]          (bu){};

    \vertex [right=2cm of b0]          (x0){};
    \vertex [below=1cm of x0]          (xd){};
    \vertex [above=1cm of x0]          (xu){};
    \propag[double, thick] (au) to (xu);
    \propag[thick]         (ad) to (xd);
    \filldraw[draw=Nonoha!75!black, fill=Nonoha!10, line width=0.45mm, rounded corners=2pt]
      (1.15,-1.15) rectangle (2.85,1.15)
      node[pos=0.5]  {\LARGE $\mathcal T$};
    \node[draw=Shino!75!black, fill=Shino!10, line width=0.32mm,
          rounded corners=1.5pt, minimum width=0.68cm, minimum height=0.54cm,
          inner sep=1.5pt] at (0.7,1) {\Large$\tau$};
    \node[draw=Shino!75!black, fill=Shino!10, line width=0.32mm,
          rounded corners=1.5pt, minimum width=0.68cm, minimum height=0.54cm,
          inner sep=1.5pt] at (3.3,1) {\Large$\tau$};
 \end{feynhand}
 \end{feynman}
\end{tikzpicture}
}
\right)
\scalebox{0.6}
{
\begin{tikzpicture}[baseline=0cm, line cap=round, line join=round]
  \begin{feynman}[baseline=(ad)]
  \begin{feynhand}
    \vertex []                         (a0){};
    \vertex [below=1cm of a0]          (ad){};
    \vertex [above=1cm of a0]          (au){};

    \vertex [right=0.7cm of a0]        (b0){};
    \vertex [below=1cm of b0]          (bd){};
    \vertex [above=1cm of b0,dot]      (bu){};

    \vertex [right=0.7cm of b0]        (x0){};
    \vertex [below=1cm of x0]          (xd){};
    \vertex [above=1cm of x0]          (xu){};
    \propag[double, thick] (au) to (bu);
    \propag[thick]         (bu) to (xu);
    \propag[thick]         (bu) to (x0);
    \propag[thick]         (ad) to (xd);
  \end{feynhand}
  \end{feynman}
\end{tikzpicture}
}
\]
    \caption{The relationship between the full three-body amplitude $\mathcal T_3$ and the reduced amplitude $\mathcal T$. Single and double lines denote stable particles and isobars, respectively, while black dots indicate the isobar dissociation vertices.}
    \label{fig:3bT}
\end{figure}
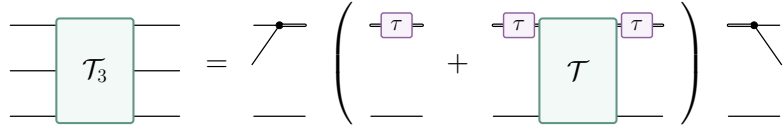

We construct the $K\bar{K}\pi$ scattering amplitude using the infinite-volume three-body unitary framework developed in Refs.~\cite{Mai:2017vot, Sadasivan:2020syi, Sadasivan:2021emk, Feng:2024wyg, Sakthivasan:2024uwd}. This approach is based on a $1+2$ decomposition, in which the full three-body amplitude $\mathcal T_3$ is expressed in terms of a reduced amplitude $\mathcal T$ describing the interaction between a spectator particle and the remaining two-body subsystem (referred to as the isobar or dimer), as illustrated in Fig.~\ref{fig:3bT}. The factors outside the parentheses correspond to the isobar dissociation vertices, which connect to the external asymptotic three-body states. Inside the parentheses, the relevant degrees of freedom are those of the spectator and the isobar, and their dynamics can be further separated into reducible and irreducible contributions. The former corresponds to processes in which the spectator does not participate in the scattering and propagates freely from the initial to the final state, effectively reducing the dynamics to a two-body scattering process. The latter encodes genuine three-body interactions and is described by the spectator-isobar amplitude $\mathcal T$. Finally, the internal dynamics of the two-body subsystem is fully accounted for by the isobar propagator, denoted by $\tau$.

Since all three-body dynamics are encoded in $\mathcal T$, we will focus exclusively on this amplitude in the following and refrain from working explicitly with $\mathcal T_3$. In particular, three-body resonance poles are contained in the analytic structure of $\mathcal T$.

Starting from the Bethe-Salpeter (BS) ansatz in scattering theory, the spectator-isobar amplitude $\mathcal T$ satisfies a BS equation. All ingredients entering this equation can be determined by imposing three-body unitarity~\cite{Mai:2017vot} together with experimental input. The resulting equation takes the form
\begin{align}\label{eq:BSE}
\mathcal T_{ji}\left(s,\boldsymbol{p}^\prime,\boldsymbol{p}\right)&=\mathcal V_{ji}\left(s,\boldsymbol{p}^\prime,\boldsymbol{p}\right)+\sumint_k\frac{\mathrm{d}^3\boldsymbol{q}}{(2\pi )^32E_{\boldsymbol{q},k}}\mathcal V_{jk}\left(s,\boldsymbol{p}^\prime,\boldsymbol{q}\right)\tau_k\left(\sigma _{\boldsymbol{q},k}\right)\mathcal T_{ki}\left(s,\boldsymbol{q},\boldsymbol{p}\right),\notag\\
\mathcal V_{ji}\left(s,\boldsymbol{p}^\prime,\boldsymbol{p}\right)&=\mathcal B_{ji}\left(s,\boldsymbol{p}^\prime,\boldsymbol{p}\right)+\mathcal C_{ji}\left(s,\boldsymbol{p}^\prime,\boldsymbol{p}\right),
\end{align}
where $i,j,k$ label the incoming, outgoing, and intermediate spectator-isobar channels, while $\boldsymbol{p}$, $\boldsymbol{p}^\prime$, and $\boldsymbol{q}$ denote the corresponding spectator momenta. The symbol $\sumint_k$ denotes a sum over the intermediate channels $k$ together with the integral over the intermediate spectator momentum $\boldsymbol{q}$, whose measure is written out explicitly. The variable $s$ is the squared invariant mass of the three-particle system (square of the total momentum $P^{(4)}$), and $\sigma_{\boldsymbol{q},k}$ is the squared invariant mass of the intermediate isobar in channel $k$:
\begin{equation}\label{eq:kin}
\sigma_{\boldsymbol{q},k}=\left(P^{(4)}-q^{(4)}\right)^2=s+m_k^2-2\sqrt{s}E_{\boldsymbol q,k}\,,\quad E_{\boldsymbol q,k}=\sqrt{\boldsymbol q^2+m_k^2}\equiv E_{q,k}\,,
\end{equation}
with $m_k$ the corresponding spectator mass. For clarity, we use $q^{(4)}$, $\boldsymbol q$, and $q$ to represent the four-momentum, the three-momentum, and the magnitude of the three-momentum, respectively.

In this BS equation, there are four building blocks to be specified. Two of them are fixed by unitarity, namely the one-particle-exchange term $\mathcal{B}$ and the isobar ``self-energy'' (encoded in $\tau$). Their on-shell contributions generate the imaginary part of the amplitude, which is fully constrained by three-body unitarity. On the other hand, unitarity does not constrain the real part of the inverse amplitude even in the physical region. Therefore, two additional off-shell inputs are required: the $\mathcal K$-matrix of the two-body subsystem (also contained in $\tau$) and a short-range three-body contact interaction $\mathcal{C}$. The former can be determined by fitting two-body scattering phase shifts, while the latter may be fixed from the lineshapes of $1\to 3$ decay processes.

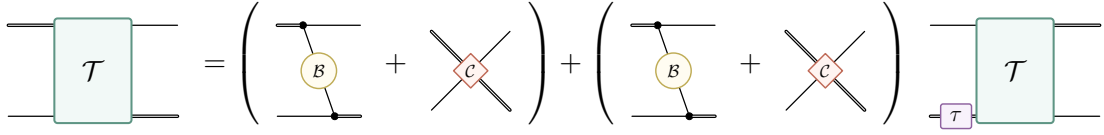
\begin{figure}[tb]
    \centering
    \[
\scalebox{0.6}
{
\begin{tikzpicture}[baseline=0cm, line cap=round, line join=round]
  \begin{feynman}[baseline=(a0d)]
  \begin{feynhand}
    \vertex []                         (a0){};
    \vertex [below=1cm of a0]          (ad){};
    \vertex [above=1cm of a0]          (au){};

    \vertex [right=2cm of a0]          (b0){};
    \vertex [below=1cm of b0]          (bd){};
    \vertex [above=1cm of b0]          (bu){};

    \vertex [right=2cm of b0]          (x0){};
    \vertex [below=1cm of x0]          (xd){};
    \vertex [above=1cm of x0]          (xu){};
    \propag[double, thick] (au) to (bu);
    \propag[double, thick] (bd) to (xd);
    \propag[thick]         (ad) to (bd);
    \propag[thick]         (bu) to (xu);
    \filldraw[draw=Nonoha!75!black, fill=Nonoha!10, line width=0.45mm, rounded corners=2pt]
      (1.15,-1.15) rectangle (2.85,1.15)
      node[pos=0.5]  {\LARGE $\mathcal T$};
 \end{feynhand}
 \end{feynman}
\end{tikzpicture}
}
=
\left(
\scalebox{0.6}
{
\begin{tikzpicture}[baseline=0cm, line cap=round, line join=round]
  \begin{feynman}[baseline=(ad)]
  \begin{feynhand}
    \vertex []                         (a0){};
    \vertex [below=1cm of a0]          (ad){};
    \vertex [above=1cm of a0]          (au){};

    \vertex [right=0.7cm of a0]        (b0){};
    \vertex [below=1cm of b0]          (bd){};
    \vertex [above=1cm of b0,dot]      (bu){};

    \vertex [right=0.7cm of b0]        (c0){};
    \vertex [below=1cm of c0,dot]      (cd){};
    \vertex [above=1cm of c0]          (cu){};

    \vertex [right=0.7cm of c0]        (x0){};
    \vertex [below=1cm of x0]          (xd){};
    \vertex [above=1cm of x0]          (xu){};
    \propag[double, thick] (au) to (bu);
    \propag[double, thick] (cd) to (xd);
    \propag[thick]         (bu) to (cd);
    \propag[thick]         (ad) to (cd);
    \propag[thick]         (bu) to (xu);
    \node[draw=Mikaru!75!black, fill=Mikaru!10, line width=0.32mm,
          circle, minimum size=0.78cm, inner sep=0pt]
          at (1.05,0) {\bf$\mathcal B$};
  \end{feynhand}
  \end{feynman}
\end{tikzpicture}
}
+
\scalebox{0.6}
{
\begin{tikzpicture}[baseline=0cm, line cap=round, line join=round]
  \begin{feynman}[baseline=(a0)]
  \begin{feynhand}
    \vertex []                         (a0){};
    \vertex [below=1cm of a0]          (ad){};
    \vertex [above=1cm of a0]          (au){};

    \vertex [right=2cm of a0]          (b0){};
    \vertex [below=1cm of b0]          (bd){};
    \vertex [above=1cm of b0]          (bu){};
    \propag[double, thick] (au) to (bd);
    \propag[thick]         (ad) to (bu);

    \node[draw=Miku!75!black, scale=1.0, diamond, fill=Miku!10,
          line width=0.32mm, inner sep=2pt] at (1,0)  {\bf$\mathcal C$};
  \end{feynhand}
  \end{feynman}
\end{tikzpicture}
}
\right)
+
\left(
\scalebox{0.6}
{
\begin{tikzpicture}[baseline=0cm, line cap=round, line join=round]
  \begin{feynman}[baseline=(ad)]
  \begin{feynhand}
    \vertex []                         (a0){};
    \vertex [below=1cm of a0]          (ad){};
    \vertex [above=1cm of a0]          (au){};

    \vertex [right=0.7cm of a0]        (b0){};
    \vertex [below=1cm of b0]          (bd){};
    \vertex [above=1cm of b0,dot]      (bu){};

    \vertex [right=0.7cm of b0]        (c0){};
    \vertex [below=1cm of c0,dot]      (cd){};
    \vertex [above=1cm of c0]          (cu){};

    \vertex [right=0.7cm of c0]        (x0){};
    \vertex [below=1cm of x0]          (xd){};
    \vertex [above=1cm of x0]          (xu){};
    \propag[double, thick] (au) to (bu);
    \propag[double, thick] (cd) to (xd);
    \propag[thick]         (bu) to (cd);
    \propag[thick]         (ad) to (cd);
    \propag[thick]         (bu) to (xu);
    \node[draw=Mikaru!75!black, fill=Mikaru!10, line width=0.32mm,
          circle, minimum size=0.78cm, inner sep=0pt]
          at (1.05,0) {\bf$\mathcal B$};
  \end{feynhand}
  \end{feynman}
\end{tikzpicture}
}
+
\scalebox{0.6}
{
\begin{tikzpicture}[baseline=0cm, line cap=round, line join=round]
  \begin{feynman}[baseline=(a0)]
  \begin{feynhand}
    \vertex []                         (a0){};
    \vertex [below=1cm of a0]          (ad){};
    \vertex [above=1cm of a0]          (au){};

    \vertex [right=2cm of a0]          (b0){};
    \vertex [below=1cm of b0]          (bd){};
    \vertex [above=1cm of b0]          (bu){};
    \propag[double, thick] (au) to (bd);
    \propag[thick]         (ad) to (bu);

    \node[draw=Miku!75!black, scale=1.0, diamond, fill=Miku!10,
          line width=0.32mm, inner sep=2pt] at (1,0)  {\bf$\mathcal C$};
  \end{feynhand}
  \end{feynman}
\end{tikzpicture}
}
\right)
\scalebox{0.6}
{
\begin{tikzpicture}[baseline=0cm, line cap=round, line join=round]
  \begin{feynman}[baseline=(a0)]
  \begin{feynhand}
    \vertex []                         (a0){};
    \vertex [below=1cm of a0]          (ad){};
    \vertex [above=1cm of a0]          (au){};

    \vertex [right=2cm of a0]          (b0){};
    \vertex [below=1cm of b0]          (bd){};
    \vertex [above=1cm of b0]          (bu){};

    \vertex [right=2cm of b0]          (x0){};
    \vertex [below=1cm of x0]          (xd){};
    \vertex [above=1cm of x0]          (xu){};
    \propag[double, thick] (ad) to (bd);
    \propag[double, thick] (bu) to (xu);
    \propag[thick]         (au) to (bu);
    \propag[thick]         (bd) to (xd);
    \filldraw[draw=Nonoha!75!black, fill=Nonoha!10, line width=0.45mm, rounded corners=2pt]
      (1.15,-1.15) rectangle (2.85,1.15)
      node[pos=0.5]  {\LARGE $\mathcal T$};
    \node[draw=Shino!75!black, fill=Shino!10, line width=0.32mm,
          rounded corners=1.5pt, minimum width=0.68cm, minimum height=0.54cm,
          inner sep=1.5pt] at (0.7,-1) {\Large$\tau$};
  \end{feynhand}
  \end{feynman}
\end{tikzpicture}
}
\]
    \caption{Graphical illustration of Eq.~\eqref{eq:BSE}. Single lines represent spectators, double lines represent isobars, the integrand in parentheses is the BS kernel that contains the one-particle exchange $\mathcal B$ and the short-range three-body force $\mathcal C$, and the fully dressed isobar propagator $\tau$ encodes the on-shell information of the two-body subsystem as well as the off-shell $\mathcal K$-matrix.}
    \label{fig:BSE}
\end{figure}

With these ingredients specified, the BS equation admits a transparent diagrammatic representation, as shown in Fig.~\ref{fig:BSE}. Throughout this work, time is taken to flow from right to left, such that the diagrammatic representation of amplitudes follows the same ordering as their operator expressions.

\subsection{Channel space}

In this subsection, we discuss the specific meaning of the channel indices appearing in Eq.~\eqref{eq:BSE}. Each channel uniquely labels a particular spectator-isobar configuration. For the $K\bar{K}\pi$ system, the spectator can be a $K$, $\bar{K}$, or $\pi$, all of which are spinless pseudoscalar mesons. Correspondingly, the isobar can be formed by the two-body combinations $K\bar{K}$, $K\pi$, and $\bar{K}\pi$ (the charge-conjugated partner of the $K\pi$ isobar).

In addition to the particle content, the isobar also carries information about the internal orbital angular momentum of the two-body subsystem. Since all three particles involved are spinless, this orbital angular momentum coincides with the effective ``spin'' of the isobar, as well as its helicity.

The spectator-isobar amplitude is then projected onto definite total quantum numbers. In principle, all spectator-isobar configurations compatible with these quantum numbers should be included. In practice, the channel space must be truncated, and this truncation constitutes one of the approximations of the calculation. Specifically, the truncation removes channels with higher isobar spins and/or higher spectator-isobar partial waves, which are expected to become important only beyond the energy region and accuracy considered here. Within a truncated channel space, the integral equation implements three-body unitarity consistently among the retained channels, including all corresponding isobar self-energies and one-particle-exchange contributions. Contributions from omitted channels, when projected to the partial waves of interest, can be regarded as part of the three-body contact terms.

\begin{table}[tb]
\caption{The spectator-isobar channels with total quantum numbers $I^G(J^{PC})=0^+(1^{++})$. The first row labels the spin and isospin combinations $(J_{\text{iso}},I_{\text{iso}})$ of the isobar; the second row presents the five channels in the helicity basis; the third row gives the four channels in the basis obtained after fixing $J=1$, where the outermost subscripts represent the orbital angular momentum between the isobar and the spectator.}
\label{tab:channels}
\center
\renewcommand{\arraystretch}{1.25}
\begin{tabular}{l|c|c|c|c}
\hline\hline
$(J_{\text{iso}},I_{\text{iso}})$
& $(0,1)$
& $(0,1/2)$
& \multicolumn{2}{c}{$(1,1/2)$}
\\
Helicity basis
& $\pi a_0$
& $(K\bar\kappa)_{+}$
& \multicolumn{2}{c}{$(K\bar K^*_{\lambda=0,\pm 1})_{+}$}
\\
$JLS$ basis
& $(\pi a_0)_P $
& $(K\bar\kappa)_{+,P}$
& $(K\bar K^*)_{+,S}$
& $(K\bar K^*)_{+,D}$
\\
\hline
\hline
\end{tabular}
\end{table}

In this work, we focus on the $I^G(J^{PC})=0^+(1^{++})$ sector. Restricting the isobar spin to $J_{\text{iso}}=0,1$, the relevant spectator-isobar channels are listed in Table~\ref{tab:channels}. The $K\bar K$ isobar can appear with $(J_{\text{iso}}, I_{\text{iso}})=(0,1)$. This channel contains the $a_0(980)$ and will be denoted by $a_0$. The $K\pi$ isobar can appear with $(J_{\text{iso}}, I_{\text{iso}})=(0,1/2)$ and $(1,1/2)$, associated with the $K_0^*(700)$ (or $\kappa$) and $K^*(892)$ channels, respectively. These labels are only shorthand for interacting two-body subsystems: the isobars are not treated as asymptotic particles. Positive and negative $G$ parity require the kaonic spectator-isobar channels to be combined as
\begin{equation}
\left(K\bar K_{\text{iso}}\right)_{+}=\frac{1}{\sqrt{2}}\left(K\bar{K}_{\text{iso}}+\bar{K}K_{\text{iso}}\right),\quad\left(K\bar K_{\text{iso}}\right)_{-}=\frac{1}{\sqrt{2}}\left(K\bar{K}_{\text{iso}}-\bar{K}K_{\text{iso}}\right),
\end{equation}
where $K_{\text{iso}}$ denotes either $\kappa$ or $K^*$. In the numerical analysis below, we neglect the $\kappa$ isobar because the $(K\bar\kappa)_{+,P}$ channel involves a relative $P$ wave and has less available phase space than the $\pi a_0$ channel. We therefore expect its contribution to be subleading in the energy region considered here. This omission removes the one-particle-exchange contributions associated with the $(K\bar\kappa)_{+,P}$ channel and should be understood as part of the channel-space truncation discussed above. The calculation consequently reduces to the coupled $\pi a_0$-$(K\bar K^*)_+$ three-body problem, within which three-body unitarity is implemented consistently.

\subsection{One-particle exchange}

We now turn to the structure of the one-particle-exchange term $\mathcal{B}$. Owing to three-body unitarity, its form is completely fixed. Since $\mathcal{B}$ and the short-range three-body interaction $\mathcal{C}$ always appear in the combination $\mathcal{V}=\mathcal{B} + \mathcal{C}$, any residual local ambiguity in $\mathcal{B}$ can be absorbed into a redefinition of $\mathcal{C}$.

The expression for $\mathcal{B}$ required by three-body unitarity takes the following form in the helicity basis:~\cite{Mai:2017vot, Sadasivan:2021emk}
\begin{equation}\label{eq:OPE}
\mathcal B_{ji}\left(\boldsymbol{p}^\prime,\boldsymbol{p}\right)=\frac{I_{ji}v_{j}^{*}\left(P^{(4)}-p^{(4)}-p^{\prime(4)},p^{(4)}\right)v_{i}\left(P^{(4)}-p^{(4)}-p^{\prime(4)},p^{\prime(4)}\right)}{2E_{\boldsymbol{p}+\boldsymbol{p}^{\prime},\text{ex}}\left( \sqrt{s}-E_{\boldsymbol{p},i}-E_{\boldsymbol{p}^\prime,j}-E_{\boldsymbol{p}+\boldsymbol{p}^\prime,\text{ex}} \right)+\ii\epsilon},
\end{equation}
where $m_i,m_j,m_\text{ex}$ form a permutation of $m_K,m_{\bar{K}},m_\pi$, with $m_i$ and $m_j$ fixed by the specific channels (incoming and outgoing spectators) and $m_\text{ex}$ the mass of the remaining particle (the exchanged one). $E_{\boldsymbol{p}+\boldsymbol{p}^\prime,\text{ex}}$ denotes the energy of the exchanged particle. The factor $I_{ji}$ is the isospin coefficient: $I_{\pi a_0\to\pi a_0}=0$, $I_{\pi a_0\to (K\bar K^*)_+}=I_{(K\bar K^*)_+\to\pi a_0}=-\sqrt 2$, and $I_{(K\bar K^*)_+\to (K\bar K^*)_+}=1$~\cite{Feng:2024wyg}. $v_i$ denotes the isobar dissociation vertex. For an $S$-wave isobar ($a_0\to K\bar K$), $v_{\lambda=0} = 1$ with momentum dependence neglected. For a $P$-wave isobar ($K^*\to K\pi$ or $\bar K^*\to \bar K\pi$), it is given by
\begin{equation}\label{eq:isobarvertex}
v_{\lambda=0,\pm1}\left(a,b\right)=g_V\epsilon_{\lambda}^\mu\left(\bm{a}+\bm b\right)\left(a^{(4)}-b^{(4)}\right)_\mu,
\end{equation}
where $a^{(4)}$ and $b^{(4)}$ denote the four-momenta of the two particles after the dissociation of the isobar, and $g_V$ is the coupling constant between the isobar and the two particles, to be determined in Sec.~\ref{sec:2b}. The helicity polarization vector of a spin-1 particle is defined as
\begin{equation}
\begin{aligned}
\epsilon_0^\mu(\bm p)=\frac{1}{m_{\text{iso}}}\begin{pmatrix}p\\E\sin\theta\cos\phi\\E\sin\theta\sin\phi\\E\cos\theta\end{pmatrix},\quad\epsilon_{\pm1}^\mu(\bm p)=\frac{1}{\sqrt{2}}\begin{pmatrix}0\\\mp\cos\theta\cos\phi+\ii\sin\phi\\\mp\cos\theta\sin\phi-\ii\cos\phi\\\pm\sin\theta\end{pmatrix},
\end{aligned}
\end{equation}
where $\boldsymbol{p}=\left(p\sin\theta\cos\phi,p\sin\theta\sin\phi,p\cos\theta\right)$ and $E=\sqrt{p^2+m_{\text{iso}}^2}$. At this point, an additional parameter $m_{\text{iso}}$ appears. This is because the construction of helicity states depends on the choice of reference frame. Accordingly, an effective ``mass'' is introduced for the isobar, which we take to be the real part of the pole position of the corresponding resonance; this quantity will also be determined in Sec.~\ref{sec:2b}.

It is worth noting that Eq.~\eqref{eq:OPE} corresponds only to the forward-propagating contribution of the one-particle-exchange diagram in time-ordered perturbation theory. If the backward-propagating contribution were also included, one would recover the standard covariant propagator $1/\left(u-m_{\text{ex}}^2+\ii\epsilon\right)$~\cite{Mai:2017vot, Zhang:2021hcl}. However, such backward-propagating terms are of short-range nature; they are not required by $K\bar K\pi$ three-body unitarity here and are therefore omitted from $\mathcal B$ (their effects can instead be absorbed into $\mathcal C$).

\subsection{Partial-wave amplitude}\label{sec:PWE}

For a function in the helicity basis, $\mathcal F_{\lambda^\prime\lambda}\left(\boldsymbol{p}^\prime,\boldsymbol{p}\right)$, where $\lambda=0$ for a scalar isobar and $\lambda=0,\pm1$ for a vector isobar, the projection onto definite total angular momentum $J$ is
\begin{equation}\label{eq:proj}
\mathcal F_{\lambda^\prime\lambda}^J\left(p^\prime,p\right)=2\pi\int_{-1}^{+1}\mathrm{d}z~\mathscr d_{\lambda^\prime\lambda}^J(z)\mathcal F_{\lambda^\prime\lambda}\left(\boldsymbol{p}^\prime,\boldsymbol{p}\right),
\end{equation}
where $\mathscr d_{\lambda^\prime\lambda}^J(z)$ denotes the Wigner $d$ function, with $z\equiv\boldsymbol{p}^\prime\cdot\boldsymbol{p}/p^\prime p$. The expression for $\mathcal F$ in the $JLS$ basis is obtained through a linear transformation:
\begin{equation}
\mathcal F_{L^\prime L}^{J=1}\left(p^\prime,p\right)=\sum_{\lambda,\lambda^\prime}U^{}_{L^\prime\lambda^\prime}\mathcal F_{\lambda^\prime\lambda}^{J=1}\left(p^\prime,p\right)U^{}_{L\lambda}\,.
\end{equation}
The transformation matrices are given by~\cite{Sadasivan:2021emk, Feng:2024wyg}
\begin{align}
&U_{L\lambda}=1~\text{for}~\pi a_0~\text{with}~L=1,~\lambda=0\,,\\
&U_{L\lambda}=\begin{pmatrix}\frac{1}{\sqrt{3}}&\frac{1}{\sqrt{3}}&\frac{1}{\sqrt{3}}\\\frac{1}{\sqrt{6}}&-\sqrt{\frac{2}{3}}&\frac{1}{\sqrt{6}}\end{pmatrix}~\text{for}~(K\bar K^*)_+~\text{with}~L=0,2,~\lambda=0,\pm1\,.
\end{align}

The BS equation in the $JLS$ basis reads as
\begin{equation}\label{eq:PWE}
\mathcal T^J_{ji}\left(s,p^\prime,p\right)=\mathcal V^J_{ji}\left(s,p^\prime,p\right)+\sumint_k\frac{q^2\mathrm{d}q}{(2\pi )^32E_{q,k}}\mathcal V^J_{jk}\left(s,p^\prime,q\right)\tau _k\left(\sigma _{q,k}\right)\mathcal T^J_{ki}\left(s,q,p\right).
\end{equation}
Now the channel indices no longer refer to the isobar helicity $\lambda$, but instead to the orbital angular momentum $L$ between the spectator and the isobar, as summarized in Table~\ref{tab:channels}. In summary, the channels considered in this work are labeled as follows: (1) $P$-wave $\pi a_0$, (2) $S$-wave $(K\bar K^*)_+$, and (3) $D$-wave $(K\bar K^*)_+$.

The integral in Eq.~\eqref{eq:PWE} is regularized by a sharp cutoff $\Lambda$. The cutoff must be large enough to cover the part of the spectator momentum range for which the intermediate isobar can decay into the two particles of the corresponding subsystem:
\begin{equation}
\sqrt{\sigma_{q,k}}\geq m_j+m_\text{ex}\,.
\end{equation}
Here $k$ refers to the intermediate channel with $m_k$ being the mass of the spectator, while $m_j$ and $m_{\rm ex}$ are the masses of the two particles forming the corresponding isobar; equivalently, $m_k,m_j,m_{\rm ex}$ form the channel-dependent permutation of $m_K,m_{\bar K},m_\pi$. Since $\sigma_{q,k}$ decreases monotonically with $q$, this condition gives the minimal cutoff needed at a given total energy:
\begin{equation}
\Lambda_{\min}=p_\text{cm}\left(\sqrt{s};\,m_k,m_j+m_\text{ex}\right),\quad p_\text{cm}\left(a;\,b,c\right)=\frac{\sqrt{\lambda\left(a^2,b^2,c^2\right)}}{2a},
\end{equation}
with $\lambda(x,y,z)=x^2+y^2+z^2-2xy-2yz-2zx$ the Källén triangle function. To cover the energy range from 1.24 to 1.60~GeV, the minimal cutoff required is found to be 0.57~GeV. Since larger cutoff values probe progressively more off-shell two-body dynamics, the choice $\Lambda=0.57$~GeV corresponds to the most conservative implementation of the three-body equations and is therefore adopted as the reference cutoff in the fits presented below. We note that alternative formulations of the three-body scattering equations, such as that of Ref.~\cite{Jackura:2018xnx}, employ an energy-dependent cutoff chosen separately for each $\sqrt{s}$, thereby restricting the input entirely to the physically accessible two-body region. Also note that one can always trade off-shell two-particle dynamics with the three-body force; see, e.g., Ref.~\cite{Polyzou:1990hks}, as mentioned above.

\section{Physical inputs}\label{sec:inputs}

\subsection{Two-body inputs}\label{sec:2b}

The three-body unitarity implemented in the formalism automatically ensures two-body unitarity~\cite{Mai:2017vot}. This constrains the imaginary part of the inverse isobar propagator $\tau^{-1}$, while leaving sufficient freedom to choose a parametrization of $\tau$ suitable for the problem at hand.

For the $\pi a_0$ channel, the $a_0$ isobar represents the $I=1$, $S$-wave $K\bar K$ subsystem. We parametrize this isobar using the coupled-channel unitarized chiral amplitude from the $N/D$ method of Ref.~\cite{Oller:1998zr}. In that work, the scalar isovector system, which is relevant here, involves two coupled two-body channels, $\eta\pi$ and $K\bar K$, which are labeled as channels $1$ and $2$, respectively. The unitarized amplitude is written as
\begin{equation}\label{eq:ollerT}
T^{I=1}_{J=0}(\sigma)=\left(\left(T^{\infty,I=1}_{J=0}(\sigma)\right)^{-1}+g_{J=0}(\sigma)\right)^{-1},
\end{equation}
where $T^{\infty,I=1}_{J=0}$ denotes the corresponding tree-level scalar amplitude and $g_{J=0}$ is a diagonal matrix of two-meson loop functions. Since the present three-body framework contains only the $K\bar K\pi$ degrees of freedom, the $\eta\pi$ channel is not included as an explicit isobar channel. We therefore identify the $a_0$ isobar propagator with the $K\bar K\to K\bar K$ matrix element of Ref.~\cite{Oller:1998zr}:
\begin{equation}\label{eq:taua0}
\tau_{a_0}(\sigma)=-\left(T^{I=1}_{J=0}(\sigma)\right)_{22},
\end{equation}
where the overall minus sign is due to the different amplitude convention. With this identification, $\tau_{a_0}$ also contains an inelastic contribution from the $\eta\pi$ channel and thus retains the width associated with $a_0\to\eta\pi$. Two versions of the scalar amplitude are provided in Ref.~\cite{Oller:1998zr}. 
Our default choice is the full model, which includes the bare resonant contributions. In this case, the $a_0$ pole is located at
\begin{equation}\label{eq:a0pole_full}
\sqrt{\sigma_{a_0}}=\left(1056-\ii\,23\right)\text{MeV}
\end{equation}
on the $(-,+)$ RS, where the first (second) sign indicates the imaginary part of the center-of-mass (c.m.) momentum of the $\eta\pi$ ($K\bar K$) channel.
Such a pole position leads to a cusp at the $K\bar K$ threshold as the shortest path for the pole to reach the physical region has to circle around the threshold~\cite{Zhang:2024qkg}. We have checked that all analytic continuations in the variable $s$ considered in this work do not cause $\sigma$ to cross the cut between the $\eta\pi$ and $K\bar{K}$ thresholds.\footnote{$\sigma$ depends on both $s$ and the spectator momentum through the kinematic relation in Eq.~\eqref{eq:kin}. Therefore, analytic continuation in $s$ induces a corresponding continuation of $\sigma$.} When searching for poles in Sec.~\ref{sec:res}, $\sigma$ may cross the cut above the $K\bar{K}$ threshold, analytically continuing $\tau_{a_0}(\sigma)$ to the $(-,-)$ RS. In this case, the corresponding branch of $\tau_{a_0}(\sigma)$ on that sheet should be used. However, the $(-,+)$ RS is never encountered. Therefore, the spectator-isobar amplitude considered in this work does not truly ``see'' the $a_0$ pole. This, in fact, also avoids several potential complications. Finally, to estimate the dependence of the resulting spectator-isobar amplitude on the underlying two-body interaction model, we also consider the reduced model in Ref.~\cite{Oller:1998zr}, in which the bare resonant contributions are removed. The corresponding pole position is then found at $(1083-\ii\,12)$~MeV on the same RS.

For the $(K\bar K^*)_+$ channel, the $K^*$ isobar propagator is dominated by a single elastic $K\pi$ channel in the energy range of interest. Owing to the simplicity of this single-channel $P$-wave scattering, a more straightforward parametrization is sufficient to satisfy the unitarity constraints. We express it in terms of a two-body $\mathcal K$-matrix and a self-energy~\cite{Sadasivan:2021emk}:
\begin{align}
\tau_{K^*}(\sigma)&=\frac{I_{K^*}^2}{\mathcal K_n^{-1}(\sigma)-\Sigma_n(\sigma)},\\
\Sigma_n(\sigma)&=\int_0^{+\infty}\frac{l^2\mathrm{d}l}{(2\pi)^3}\frac{E_{l,K}+E_{l,\pi}}{2E_{l,K}E_{l,\pi}}\left(\frac{\sigma}{\left(E_{l,K}+E_{l,\pi}\right)^2}\right)^n\frac{\tilde{v}^2(l)}{\sigma-\left(E_{l,K}+E_{l,\pi}\right)^2+\ii\epsilon}.
\end{align}
Here, $n$ denotes the number of subtractions, and ${I_{K^*}}=\sqrt{3/4}$ is the isospin factor~\cite{Feng:2024wyg}. The quantity $\tilde{v}$ again represents the isobar dissociation vertex, now projected onto the spin quantum numbers of the isobar. For a $P$-wave isobar, it is given by
\begin{equation}
\tilde{v}(l)={I_{K^*}}\sqrt{\frac{16\pi}{3}}g_Vl\,,
\end{equation}
where $g_V$ is the coupling constant between the isobar and the two particles. The $n$-times subtracted $\mathcal K$-matrix is parametrized as $\mathcal{K}_n^{-1}(\sigma)=\sum_{m=0}^{n-1} a_m\sigma^m$. For $n=2$, choosing $a_0 = -m_V^2$ and $a_1 = 1$, the isobar propagator $\tau$ reduces to the standard form:
\begin{equation}
\tau_{K^*}(\sigma)=\frac{{I_{K^*}^2}}{\sigma - m_V^2 - \Sigma_2(\sigma)},
\end{equation}
where $m_V$ can be interpreted as the ``bare mass'' of the isobar. This particular parametrization for a $P$-wave isobar already provides a sufficiently accurate description of the physical on-shell two-body amplitude. The parameters $g_V$ and $m_V$ are taken from Ref.~\cite{Feng:2024wyg}, where they were determined from fits to the experimental $K\pi$ $P$-wave scattering phase shifts. The corresponding partial-wave amplitude and phase shift are given by
\begin{equation}\label{eq:2bT}
T^{I=1/2}_{J=1}(\sigma)=\frac{\tilde{v}^2\left(p_\text{cm}(\sqrt \sigma;\,m_1,m_2)\right)}{\sigma - m_V^2 - \Sigma_2(\sigma)},\quad\delta^{I=1/2}_{J=1}(\sigma)=\arctan\left(\frac{\mathrm{Im}~T^{I=1/2}_{J=1}(\sigma)}{\mathrm{Re}~T^{I=1/2}_{J=1}(\sigma)}\right).
\end{equation}
The resulting $K^*$ pole position is located on the second RS at
\begin{equation}\label{eq:K*pole}
\sqrt{\sigma_{K^*}}=\left(899-\ii\,29\right)\text{MeV}\,.
\end{equation}
For a review on dispersive analyses of the $\pi K$ scattering amplitude, see Ref.~\cite{Pelaez:2021dak}.

\subsection{Three-body inputs}

\begin{figure}[tb]
    \centering
    \[
\begin{tikzpicture}[baseline=0cm, line cap=round, line join=round]
\begin{feynman}[baseline=(a0)]
\begin{feynhand}
\vertex                                (xx){};
\vertex [above=1cm of xx]              (x1){};
\vertex [below=1cm of xx]              (x2){};
\vertex [above=1cm of x1]              (x3){};

\vertex [right=1.9cm of xx]            (a0){};
\vertex [above=0.065cm of a0]          (au){};
\vertex [below=0.065cm of a0]          (ad){};

\vertex [right=2.55cm of a0]           (b0){};
\vertex [above=0.065cm of b0]          (bu){};
\vertex [below=0.065cm of b0]          (bd){};

\vertex [right=1.5cm of b0]            (c0){};

\propag[sca] (b0)   to [edge label={$J/\psi$}]   (c0);
\propag[thick] (au)   to [edge label={$f_1$}]      (bu);
\propag[thick] (a0)   to                           (b0);
\propag[thick] (ad)   to                           (bd);
\propag[thick] (xx)   to                           (a0);
\propag[thick] (x1)   to [out=0, in=120]           (a0);
\propag[thick] (x2)   to [out=0, in=-120]          (a0);
\propag[pho] (x3)   to [out=0, in=120] [edge label={$\gamma$}] (b0);

\node[draw=Toko!75!black, fill=Toko!10, line width=0.32mm,
      rounded corners=2pt, minimum width=0.78cm, minimum height=0.56cm,
      inner sep=0pt] at (1.9,0)  {\bf$\mathcal A$};
\filldraw[fill=black, line width=0.2mm] (4.45,0) circle [radius=0.15cm];

\end{feynhand}
\end{feynman}
\end{tikzpicture}
\]
    \caption{Graphical illustration of the $J/\psi\to\gamma(K^0_SK^0_S\pi^0)$ decay.}
    \label{fig:4decay}
\end{figure}
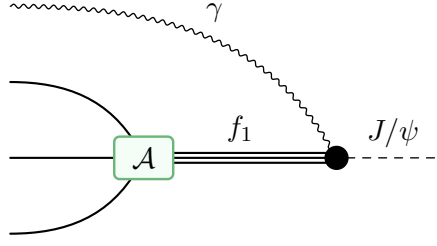

The short-range three-body interaction $\mathcal{C}$, or equivalently the three-body $\mathcal K$ matrix, para\-me\-trizes the part of the real amplitude that is not fixed by three-body unitarity. Its low-energy constants absorb the effects of degrees of freedom outside the explicit $K\bar K\pi$ channel space and depend on the ultraviolet cutoff $\Lambda$. They must therefore be determined from external input. Here we fit them to the $0^+(1^{++})$ contribution to the $K^0_SK^0_S\pi^0$ lineshape extracted by the BESIII partial-wave analysis of the process $J/\psi\to\gamma(K^0_SK^0_S\pi^0)$~\cite{BESIII:2022chl}, depicted in Fig.~\ref{fig:4decay}.

In this work, the three-body force is parametrized as a spectator-isobar contact potential in the channel basis of Sec.~\ref{sec:PWE}:
\begin{equation}\label{eq:C}
\mathcal{C}_{ji}^J\left( s,p^\prime ,p \right) =\left( \begin{matrix}
\left( c_{11}+\frac{h_1^2m_{\pi}^{2}}{s-m_{f}^{2}} \right) \frac{p^\prime p}{m_{\pi}^{2}}&		\left( c_{12}+\frac{h_1h_2m_{\pi}^{2}}{s-m_{f}^{2}} \right) \frac{p^\prime}{m_{\pi}}&		0\\
\left( c_{12}+\frac{h_1h_2m_{\pi}^{2}}{s-m_{f}^{2}} \right) \frac{p}{m_{\pi}}&		c_{22}+\frac{h_2^2m_{\pi}^{2}}{s-m_{f}^{2}}&		0\\
0&		0&		0\\
\end{matrix} \right).
\end{equation}
Since the $f_1(1285)$ lies somewhat far from the relevant thresholds, it may not be generated dynamically by constant interactions alone. We therefore explicitly introduce a bare state with a bare mass $m_f$, which represents an effective degree of freedom of the Castillejo-Dalitz-Dyson-pole type rather than a state with a definite nonstrange or strange quark-model assignment. It should be understood as an effective linear combination of unresolved compact configurations, not as a state with a model-independent flavor wave function. In Sec.~\ref{sec:res}, we will examine how this bare state, once dressed, evolves into a physical pole.

\begin{figure}[tb]
    \centering
    \[
        \scalebox{0.6}{
        \begin{tikzpicture}[baseline=0cm, line cap=round, line join=round]
        \begin{feynman}[baseline=(c1)]
        \begin{feynhand}
        \vertex                                    (c1){};
        \vertex [left=1.9cm of c1]                 (d1){};
        \vertex [above=1cm of d1]                  (d2){};
        \vertex [below=1cm of d1]                  (d3){};

        \propag[thick] (c1) to                  (d1);
        \propag[thick] (c1) to [out=120, in=0]  (d2);
        \propag[thick] (c1) to [out=-120, in=0] (d3);

        \node[draw=Toko!75!black, fill=Toko!10, line width=0.32mm,
              rounded corners=2pt, minimum width=0.78cm, minimum height=0.56cm,
              inner sep=0pt] at (0.0,0)  {\bf$\mathcal A$};

        \end{feynhand}
        \end{feynman}
        \end{tikzpicture}
        }
        =
        \scalebox{0.6}
        {
        \begin{tikzpicture}[baseline=0cm, line cap=round, line join=round]
        \begin{feynman}[baseline=(ad)]
        \begin{feynhand}
        \vertex []                         (a0){};
        \vertex [below=1cm of a0]          (ad){};
        \vertex [above=1cm of a0]          (au){};

        \vertex [left=0.7cm of a0]         (b0){};
        \vertex [below=1cm of b0]          (bd){};
        \vertex [above=1cm of b0,dot]      (bu){};

        \vertex [left=0.7cm of b0]         (x0) {};
        \vertex [below=1cm of x0]          (xd) {};
        \vertex [above=1cm of x0]          (xu) {};

        \propag[double, thick] (au) to (bu);
        \propag[thick]         (bu) to (xu);
        \propag[thick]         (bu) to (x0);
        \propag[thick]         (ad) to (xd);
        \end{feynhand}
        \end{feynman}
        \end{tikzpicture}
        }
        \left(
        \scalebox{0.6}
        {
        \begin{tikzpicture}[baseline=0cm, line cap=round, line join=round]
        \begin{feynman}[baseline=(a0)]
        \begin{feynhand}
        \vertex []                         (a0){};
        \vertex [below=1cm of a0]          (ad){};
        \vertex [above=1cm of a0]          (au){};

        \vertex [right=0.7cm of a0]        (b0){};
        \vertex [below=1cm of b0]          (bd){};
        \vertex [above=1cm of b0,dot]      (bu){};

        \vertex [right=0.7cm of b0]        (x0){};
        \vertex [below=1cm of x0]          (xd){};
        \vertex [above=1cm of x0]          (xu){};
        \propag[double, thick] (au) to (xu);
        \propag[thick]         (ad) to (xd);
        \node[draw=Shino!75!black, fill=Shino!10, line width=0.32mm,
              rounded corners=1.5pt, minimum width=0.68cm, minimum height=0.54cm,
              inner sep=1.5pt] at (0.7,1) {\Large$\tau$};
        \end{feynhand}
        \end{feynman}
        \end{tikzpicture}
        }
        +
        \scalebox{0.6}
        {
        \begin{tikzpicture}[baseline=0cm, line cap=round, line join=round]
        \begin{feynman}[baseline=(a0d)]
        \begin{feynhand}
        \vertex []                         (a0){};
        \vertex [below=1cm of a0]          (ad){};
        \vertex [above=1cm of a0]          (au){};

        \vertex [right=2cm of a0]          (b0){};
        \vertex [below=1cm of b0]          (bd){};
        \vertex [above=1cm of b0]          (bu){};

        \vertex [right=2cm of b0]          (x0){};
        \vertex [below=1cm of x0]          (xd){};
        \vertex [above=1cm of x0]          (xu){};
        \propag[double, thick] (au) to (xu);
        \propag[thick]         (ad) to (xd);
        \filldraw[draw=Nonoha!75!black, fill=Nonoha!10, line width=0.45mm, rounded corners=2pt]
          (1.15,-1.15) rectangle (2.85,1.15)
          node[pos=0.5]  {\LARGE $\mathcal T$};
        \node[draw=Shino!75!black, fill=Shino!10, line width=0.32mm,
              rounded corners=1.5pt, minimum width=0.68cm, minimum height=0.54cm,
              inner sep=1.5pt] at (0.7,1) {\Large$\tau$};
        \node[draw=Shino!75!black, fill=Shino!10, line width=0.32mm,
              rounded corners=1.5pt, minimum width=0.68cm, minimum height=0.54cm,
              inner sep=1.5pt] at (3.3,1) {\Large$\tau$};
        \end{feynhand}
        \end{feynman}
        \end{tikzpicture}
        }
        \right)
        \scalebox{0.6}{
        \begin{tikzpicture}[baseline=0cm, line cap=round, line join=round]
        \begin{feynman}[baseline=(c1)]
        \begin{feynhand}
        \vertex                                    (c1){};
        \vertex [left=1.9cm of c1]                 (d1){};
        \vertex [above=1cm of d1]                  (d2){};
        \vertex [below=1cm of d1]                  (d3){};

        \propag[double, thick] (c1) to [out=120, in=0]  (d2);
        \propag[thick]         (c1) to [out=-120, in=0] (d3);

        \node[draw=Chima!75!black, fill=Chima!10, line width=0.32mm,
              rounded corners=2pt, minimum width=0.78cm, minimum height=0.56cm,
              inner sep=0pt] at (0.0,0)  {\bf$\mathcal D$};
        \end{feynhand}
        \end{feynman}
        \end{tikzpicture}
        }
\]
    \caption{Graphical illustration of the $0^+(1^{++})\to K\bar K\pi$ decay.}
    \label{fig:3decay}
\end{figure}
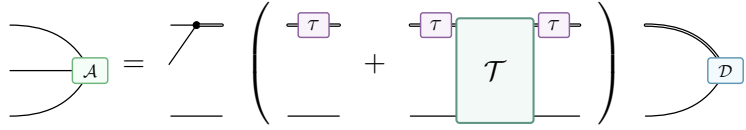

The decay process is described by first creating a spectator-isobar pair and subsequently resumming the final-state interaction through the same unitary amplitude $\mathcal T$ that appears in Eq.~\eqref{eq:PWE}; see Fig.~\ref{fig:3decay}. In the $JLS$ basis, the decay amplitudes into the $\pi a_0$ and $(K\bar K^*)_+$ channels are written as
\begin{equation}\label{eq:prod}
\mathcal A^J_{j}\left(s,p^\prime\right)=\tau_j\left(\sigma_{p^\prime,j}\right)\left(\mathcal D^J_{j}\left(p^\prime\right)+\sumint_i\frac{p^2\mathrm{d}p}{(2\pi )^32E_{p,i}}\mathcal T_{ji}^J\left(s,p^\prime,p\right)\tau_i\left(\sigma_{p,i}\right)\mathcal D^J_{i}(p)\right),
\end{equation}
where
\begin{equation}\label{eq:D}
\mathcal D_i^J(p)=\left( \begin{matrix}
	\left( d_1+\frac{d_fh_1m_{\pi}^{2}}{s-m_{f}^{2}} \right) \frac{p}{m_{\pi}}~~&		d_2+\frac{d_fh_2m_{\pi}^{2}}{s-m_{f}^{2}}~~&		0\\
\end{matrix} \right).
\end{equation}
After substituting Eq.~\eqref{eq:PWE} into Eq.~\eqref{eq:prod}, the decay amplitude can equivalently be obtained from the integral equation
\begin{equation}\label{eq:prod2}
\mathcal A^J_{j}\left(s,p^\prime\right)=\tau_j\left(\sigma_{p^\prime,j}\right)\left(\mathcal D^J_{j}\left(p^\prime\right)+\sumint_{i}\frac{p^2\mathrm{d}p}{(2\pi )^32E_{p,i}}\mathcal V_{ji}^J\left(s,p^\prime,p\right)\mathcal A^J_{i}\left(s,p\right)\right).
\end{equation}

The phase-space integral below requires the decay amplitudes $\mathcal A^J_j(s,p^\prime)$ for real spectator momenta $p^\prime$. In the numerical lineshape calculation, we do not determine these real-axis amplitudes by constructing a separate contour for each value of $p^\prime$ (see, e.g., Appendix~\ref{app:CD}). Instead, we first solve Eq.~\eqref{eq:prod2} on a fixed complex contour and then continue the result back to the real axis using the AAA algorithm~\cite{Nakatsukasa:2018aaa}.\footnote{The AAA algorithm is used only to continue the spectator-momentum dependence of the decay amplitude back to real momenta for the physical lineshape. The resonance poles reported in this work are poles in the complex $s$ plane and are instead obtained by solving the integral equation on a deformed spectator-momentum contour, and will not be affected by spurious poles of the AAA rational approximant.}

The fixed contour used for this purpose is the baseline spectator momentum contour (bSMC):~\cite{Sadasivan:2021emk} 
\begin{equation}\label{eq:bSMC}
\text{bSMC}=\left\{\,t+\ii V_0\left(1-e^{-t/w}\right)\left(1-e^{(t-\Lambda)/w}\right)~\middle|~t\in\left[0,\Lambda\right]\right\},
\end{equation}
where the real variable $t$ parametrizes the contour. Its two endpoints reproduce those of the original real-momentum integration in Eq.~\eqref{eq:prod2}: the factors $1-e^{-t/w}$ and $1-e^{(t-\Lambda)/w}$ vanish at $t=0$ and $t=\Lambda$, respectively, so the contour runs from $0$ (at $t=0$) to the real cutoff $\Lambda$ (at $t=\Lambda$) and dips into the lower half-plane in between. The parameter $w>0$ sets the slope with which the contour leaves and rejoins the real axis at the two endpoints, while $V_0$ fixes the scale of the excursion away from the real axis; a deformation into the lower half-plane corresponds to $V_0<0$, and a larger $|V_0|$ reaches further into the complex plane~\cite{Sadasivan:2021emk}. These two parameters are chosen such that the bSMC remains a one-sided deformation into the lower half-plane that avoids the three-body cuts (condition~2 in Appendix~\ref{app:CD}). We use a slight deformation for the baseline contour employed in the fits ($w=0.2$~GeV, $V_0=-0.1$~GeV), and a larger $|V_0|$ when a deeper analytic continuation is required for the pole search in Sec.~\ref{sec:res} ($w=0.2$~GeV, $V_0=-0.3$~GeV). For the quadrature nodes $p^\prime\in\mathrm{bSMC}$, the decay amplitude is obtained directly by inverting Eq.~\eqref{eq:prod2}:
\begin{equation}\label{eq:prod3}
\mathcal A^J_{j}\left(s,p^\prime\right)=\sumint_{k,q\in\text{bSMC}}\mathrm d q\left(1-\tau\mathcal V^J \right)^{-1}_{jk}\left(s,p^\prime,q\right)\tau_k\left(\sigma_{q,k}\right)\mathcal D^J_{k}(q)\,.
\end{equation}
For each fixed $s$ and each channel $j$, the computed values $\mathcal A^J_j(s,p^\prime)$ on the bSMC are continued to the real axis with the AAA algorithm~\cite{Nakatsukasa:2018aaa}.\footnote{We have also continued the same amplitudes with the rational analytic continuation method of Ref.~\cite{Sakthivasan:2024uwd}---a Thiele/Schlessinger continued-fraction interpolation~\cite{Abramowitz:1964, Schlessinger:1968vsk} of the rational-approximation family used in hadron physics~\cite{Tripolt:2016cya, Binosi:2019ecz}---and obtain consistent results.} 

The AAA algorithm represents the amplitude by a rational approximant $\mathcal R^J_j(s,z)$, with $z$ the complex spectator momentum, in the barycentric form:
\begin{equation}\label{eq:AAA}
\mathcal R^J_j(s,z)=\left(\sum_{n=1}^N\frac{w_n\,\mathcal A^J_j(s,z_n)}{z-z_n}\right)\Bigg/\left(\sum_{n=1}^N\frac{w_n}{z-z_n}\right),
\end{equation}
where the support points $\{z_n\}$ are a subset of the bSMC nodes and $\{w_n\}$ are the barycentric weights; by construction $\mathcal R^J_j(s,z_n)=\mathcal A^J_j(s,z_n)$ at the support points. AAA selects the support points iteratively: starting from the mean of the data, at each step it selects as a support point the node at which the current approximant has the largest residual $|\mathcal A^J_j-\mathcal R^J_j|$. For a given support set, the weights are fixed by the remaining (``test'') nodes $\{z_m\}$ through the Loewner matrix
\begin{equation}\label{eq:AAALoewner}
\mathbb L_{mn}=\frac{\mathcal A^J_j(s,z_m)-\mathcal A^J_j(s,z_n)}{z_m-z_n}.
\end{equation}
The barycentric weights $w_n$ are the components of a vector $\boldsymbol w=(w_1,\dots,w_N)$. Since $\mathcal R^J_j$ is invariant under a common rescaling $w_n\to c\,w_n$, the vector $\boldsymbol w$ is fixed only up to normalization, and AAA takes it to be the unit vector minimizing $\|\mathbb L\,\boldsymbol w\|$, i.e., the right singular vector of $\mathbb L$ with the smallest singular value. The order $N$ is increased until the largest residual on the test nodes drops below a chosen tolerance, and a final cleanup removes spurious Froissart poles, which carry a negligible residue or sit next to a canceling zero, so that the continuation is stable against small changes of the input data. The real-axis amplitude used in the lineshape is then
\begin{equation}
\mathcal A^J_j\left(s,p^\prime\right)\simeq \mathcal R^J_j\left(s,p^\prime\right),\quad p^\prime\in\mathbb R\,.
\end{equation}
Practically, Eq.~\eqref{eq:prod3} is solved once on the bSMC for each value of $s$, and the AAA approximant is then fit once per channel and evaluated at the real spectator momenta encountered in the three-body phase-space integration. This is the method used in the fits because it is much faster than constructing target-dependent deformed contours, and because the barycentric form is cheap to evaluate at every momentum. Appendix~\ref{app:CD} describes the corresponding contour-deformation construction, which provides a rigorous way to obtain the same real-axis amplitudes.

To construct the physical $K\bar K\pi$ decay amplitude, we transform the spectator-isobar amplitudes back to the helicity basis and attach the appropriate isobar decay vertices~\cite{Sadasivan:2020syi, Sadasivan:2021emk}. The $\pi a_0$ contribution reads as
\begin{equation}
\mathcal A^J_{\pi a_0,\lambda\Lambda}\left(s,\boldsymbol{p}_1,\boldsymbol{p}_2,\boldsymbol{p}_3\right)=\sqrt{\frac{3}{4\pi}}\mathcal A^J_{\pi a_0}\left(s,p_1\right)\delta_{0\lambda}\left(\mathscr{D}_{\lambda\Lambda}^{1}\right)^*\!\left(0,\theta_1,\phi_1\right),
\end{equation}
while the $(K\bar K^*)_+$ contribution is
\begin{equation}
\mathcal A^J_{(K\bar K^*)_+,\lambda\Lambda}\left(s,\boldsymbol{p}_1,\boldsymbol{p}_2,\boldsymbol{p}_3\right)=v_\lambda\left(\boldsymbol{p}_2,\boldsymbol{p}_3\right)\sum_L\sqrt{\frac{3}{4\pi}}\mathcal A^J_{(K\bar K^*)_+,L}\left(s,p_1\right)U_{L\lambda}\left(\mathscr{D}_{\lambda\Lambda}^{1}\right)^*\!\left(0,\theta_1,\phi_1\right),
\end{equation}
with $\mathscr{D}_{\lambda\Lambda}^{J=1}(0,\theta_1,\phi_1)$ the Wigner $D$ function~\cite{Jacob:1959at, Berman:1965gi, Chung:1971ri}. $\theta_1$ and $\phi_1$ give the polar and azimuthal angles, respectively, of $\bm p_1$. Combining the isospin coefficients and symmetrizing the two neutral kaons gives the amplitude for $0^+(1^{++})\to K^0_S(\boldsymbol{p}_1)K^0_S(\boldsymbol{p}_2)\pi^0(\boldsymbol{p}_3)$:
\begin{align}\label{eq:finalA}
\mathcal A^J_{\lambda\Lambda}\left(s,\boldsymbol{p}_1,\boldsymbol{p}_2,\boldsymbol{p}_3\right)=&-\frac{1}{\sqrt 6}\mathcal A^J_{\pi a_0,\lambda\Lambda}\left(s,\boldsymbol{p}_3,\boldsymbol{p}_1,\boldsymbol{p}_2\right)\notag\\
&-\frac{1}{2\sqrt 3}\left(\mathcal A^J_{(K\bar K^*)_+,\lambda\Lambda}\left(s,\boldsymbol{p}_1,\boldsymbol{p}_2,\boldsymbol{p}_3\right)+\left(\boldsymbol{p}_1\leftrightarrow\boldsymbol{p}_2\right)\right).
\end{align}

Finally, the invariant-mass distribution used in the fit is
\begin{equation}
\frac{\mathrm d \Gamma\left(\sqrt s\right)}{\mathrm d \sqrt s}=\sqrt s\left(m_{J/\psi}^2-s\right)\int\mathrm{d}\Phi _3\left(s,\boldsymbol{p}_1,\boldsymbol{p}_2,\boldsymbol{p}_3\right)\sum_\Lambda w_\Lambda\left|\sum_{\lambda}\mathcal A_{\lambda\Lambda}\left(s,\boldsymbol{p}_1,\boldsymbol{p}_2,\boldsymbol{p}_3\right)\right|^{2},
\end{equation}
with weights $w_0=1/2$ and $w_{\pm1}=1/4$~\cite{Nakamura:2023hbt}. The three-body phase space measure is
\begin{align}
\mathrm{d}\Phi _3\left(s,\boldsymbol{p}_1,\boldsymbol{p}_2,\boldsymbol{p}_3\right)=&\,\frac{\mathrm{d}^{3}\boldsymbol p_{1}}{(2\pi)^{3}2E_{\boldsymbol{p}_{1}}}\frac{\mathrm{d}^{3}\boldsymbol p_{2}}{(2\pi)^{3}2E_{\boldsymbol{p}_{2}}}\frac{\mathrm{d}^{3}\boldsymbol p_{3}}{(2\pi)^{3}2E_{\boldsymbol{p}_{3}}}(2\pi)^{4}\delta^{(4)}\left(P^{(4)}-p^{(4)}_{1}-p^{(4)}_{2}-p^{(4)}_{3}\right)\notag\\
=&\,\frac{p_1\bar p_{3}}{(2\pi)^58\sqrt{s}}\mathrm{d}\sqrt{s_{23}}\mathrm{d}\Omega _1\mathrm{d}\bar\Omega _{3}\,,
\end{align}
where $\bar p_{3}$ and $\bar\Omega _{3}$ are the momentum and solid angle of particle 3 in the c.m. frame of particles 2 and 3.

\section{Results}\label{sec:res}

\begin{table}[tb]
\caption{Fit parameters and pole positions for the three schemes used in this work. The first column is the reference fit. The quoted errors in this column are statistical uncertainties from the fit. The second and third columns are used to estimate the model dependence associated with the cutoff and the $K\bar K$ two-body input, respectively.}
\label{tab:res}
\center
\renewcommand{\arraystretch}{1.25}
\begin{tabular}{lccc}
\hline\hline
$K\bar K$ model
& full
& full
& reduced
\\
$\Lambda$ [GeV]
& $0.57$
& $0.73$
& $0.57$
\\
\hline
$c_{11}$
& $-11.4^{+1.6}_{-1.0}$
& $-11.2$
& $-12.1$
\\
$c_{12}$
& $44^{+12}_{-11}$
& $44$
& $48$
\\
$c_{22}$
& $-2570\pm280$
& $-1689$
& $-2577$
\\
$h_{1}$
& $7.0\pm0.3$
& $8.5$
& $7.1$
\\
$h_{2}$
& $-85^{+7}_{-4}$
& $-53$
& $-84$
\\
$m_{f}$ [MeV]
& $1359^{+16}_{-14}$
& $1350$
& $1361$
\\
$d_{1}$ [arbitrary units]
& $38^{+6}_{-7}$
& $33$
& $38$
\\
$d_{2}$ [arbitrary units]
& $195^{+70}_{-82}$
& $257$
& $181$
\\
$d_f$ [arbitrary units]
& $74^{+6}_{-4}$
& $89$
& $75$
\\
\hline
$\chi^2/\text{dof}$
& $1.9$
& $1.5$
& $1.9$
\\
$f_{1}(1285)$ pole [MeV]
& $1277\pm2-\ii\,(12\pm1)$
& $1278-\ii\,12$
& $1277-\ii\,12$
\\
$f_{1}(1420)$ pole [MeV]
& $1435\pm2-\ii\,(40\pm2)$
& $1442-\ii\,41$
& $1435-\ii\,40$
\\
Extra pole [MeV]
& $1363^{+19}_{-17}-\ii\,(106^{+12}_{-9})$
& $1409-\ii\,61$
& $1371-\ii\,103$
\\
\hline
\hline
\end{tabular}
\end{table}

\begin{figure}[tb]
    \centering
    \includegraphics[width=0.75\textwidth]{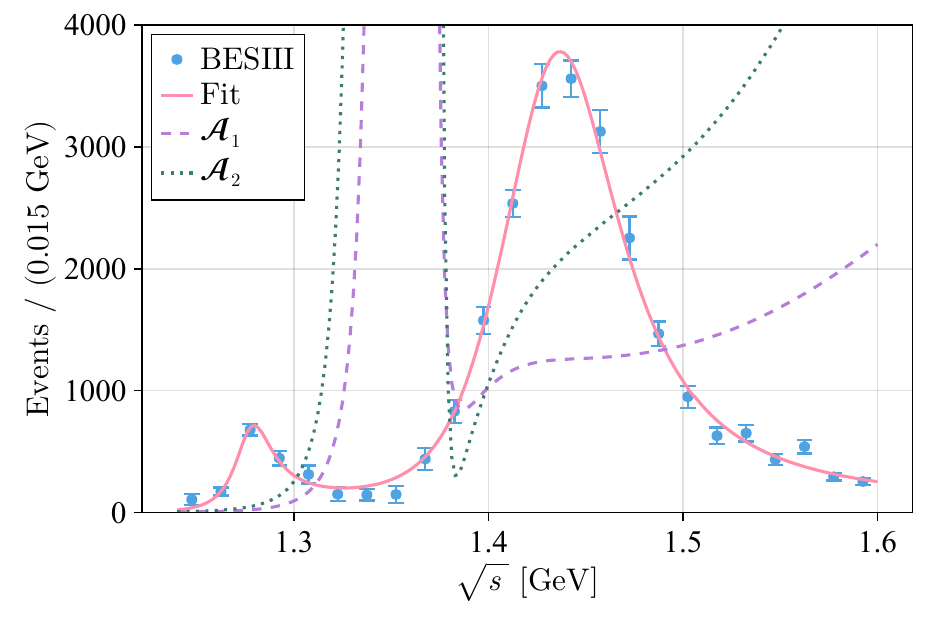}
    \caption{Fit to the $I^G(J^{PC})=0^+(1^{++})$ partial wave from the BESIII analysis of the $K^0_SK^0_S\pi^0$ invariant-mass distribution in the decay $J/\psi\to\gamma(K^0_SK^0_S\pi^0)$~\cite{BESIII:2022chl}. The solid curve shows the full result obtained from the resummed spectator-isobar amplitude. The dashed (dotted) curve shows the contribution obtained by truncating the decay amplitude after the first (second) rescattering term, i.e., $\mathcal A_1$ ($\mathcal A_2$) with $\mathcal A_N=\tau\sum_{n=0}^{N}(\mathcal V\tau)^n\mathcal D$. The dashed curve ($\mathcal A_1$) contains the bare-state contribution and the single triangle-diagram mechanism, while the dotted curve ($\mathcal A_2$) additionally includes the second-order (two-loop) rescattering contribution.}
    \label{fig:fit_lineshape}
\end{figure}

The parameters in Eqs.~\eqref{eq:C} and~\eqref{eq:D} were determined by fitting to the $0^+(1^{++})$ partial wave of the $ K^0_SK^0_S\pi^0$ invariant-mass distribution from the partial-wave analysis by the BESIII Collaboration~\cite{BESIII:2022chl}. We considered three fit schemes, summarized in Table~\ref{tab:res}. The first column is our reference fit: it uses the full $K\bar K$ scalar amplitude of Ref.~\cite{Oller:1998zr} and the cutoff $\Lambda=0.57$~GeV. The second fit keeps the same two-body input but increases the cutoff to $\Lambda=0.73$~GeV, while the third fit returns to $\Lambda=0.57$~GeV and replaces the full $K\bar K$ scalar amplitude with the reduced model discussed in Sec.~\ref{sec:2b}. The latter two fits are used to estimate the model dependence associated with the cutoff and the two-body input. All three schemes describe the measured lineshape well, with a $\chi^2/\mathrm{dof}$ in the range of 1.5--1.9. The fact that $\chi^2/\mathrm{dof}$ does not reach values very close to unity is mainly due to the valley region and one point around 1.56~GeV sticking out of the curve. Nevertheless, the overall agreement between the fitted curves and the data is good. The fitted contact parameters vary between schemes, as expected for cutoff-dependent short-range interactions.

In Fig.~\ref{fig:fit_lineshape}, the reference fit is compared with the BESIII data~\cite{BESIII:2022chl}. The fully resummed amplitude (solid curve) describes the data well, reproducing both the narrow $f_1(1285)$ peak and the pronounced peak around $1.43$~GeV. To gain a clear picture of the origin of the pronounced peak, we also show the curves obtained using the truncated amplitudes $\mathcal A_1$ (dashed) and $\mathcal A_2$ (dotted), defined as $\mathcal A_N=\tau\sum_{n=0}^{N}(\mathcal V\tau)^n\mathcal D$, which contain the bare-state contribution together with the single triangle diagram and, for $\mathcal A_2$, the two-loop rescattering correction. Both truncations diverge near the bare mass $m_f\approx1.36$~GeV, reflecting the lack of renormalization of the bare state. They show an enhancement above 1.4~GeV originating from a triangle diagram with a triangle singularity, but neither of them reproduces the peak near $1.43$~GeV. Moreover, the sizable difference between $\mathcal A_1$ and $\mathcal A_2$ in this region shows that the rescattering expansion does not converge there.\footnote{This is different from the case discussed in Ref.~\cite{Sakthivasan:2024uwd}, where no nearby pole exists and the rescattering series was found to converge rapidly, with the leading triangle diagram almost coinciding with the fully resummed result, so that the triangle singularity is only mildly affected by the final-state interactions.} The peak around $1.43$~GeV thus cannot be attributed to a single triangle diagram containing the triangle singularity. The spectator-isobar interaction is nonperturbative, and the infinite-order resummation is indispensable. The resummed amplitude indeed contains a pole corresponding to the $f_1(1420)$, as the pole analysis below demonstrates.

\begin{figure}[tb]
    \centering
    \includegraphics[width=0.75\textwidth]{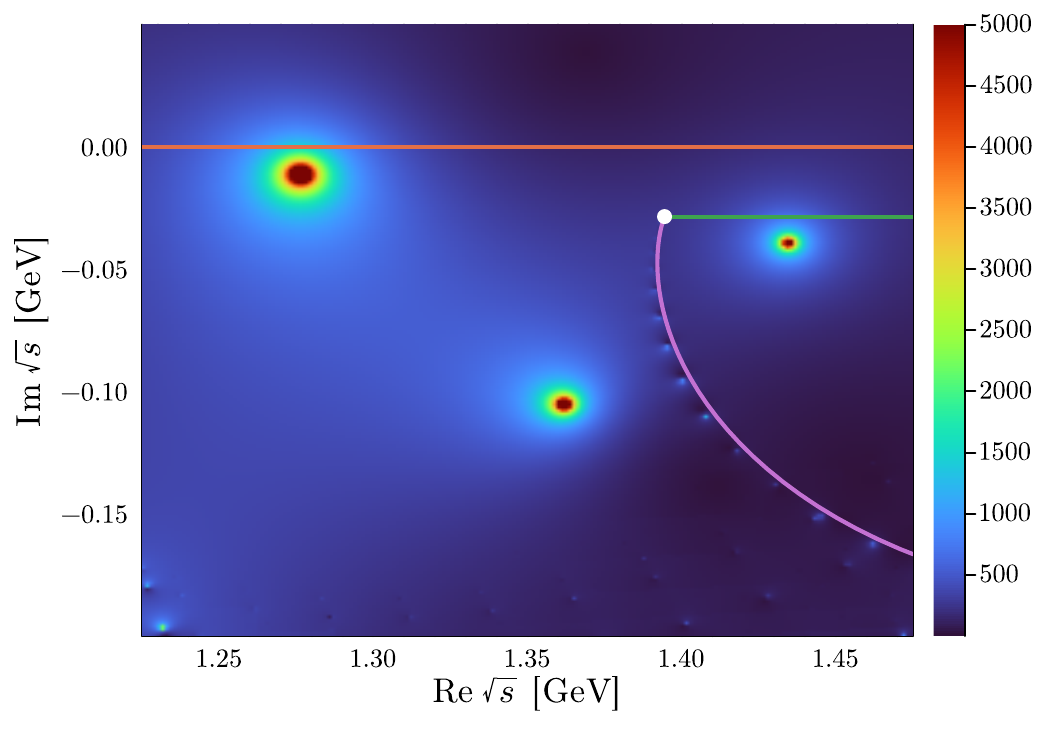}
    \caption{Density plot of the absolute value of a typical amplitude $|\mathcal{T}|$, obtained from the reference fit in Table~\ref{tab:res}, in the complex $\sqrt{s}$ plane. In addition to the two poles associated with the $f_1(1285)$ and $f_1(1420)$, an extra pole is found deeper in the complex plane on the same RS as the $f_1(1285)$. The orange and green lines represent the $K\bar{K}\pi$ and $(K\bar K^*)_+$ right-hand cuts, respectively. The white dot marks the $(K\bar K^*)_+$ branch point: $\sqrt{s}=m_K+\sqrt{\sigma_{K^*}}$. The magenta curve shows the mapping of the integration contour onto the $\sqrt{s}$ complex plane via Eq.~\eqref{eq:kin} and $\sigma=\sigma_{K^*}$; this curve coincides with a series of singularities introduced by the discretization of the integration contour. For $\operatorname{Im}\sqrt{s}<-0.15$~GeV, the amplitude shows some numerical instabilities because this region is not accessible with the present integration contour, which is also indicated by dense vertical hatching in Fig.~\ref{fig:trajectory}.}
    \label{fig:pole_positions}
\end{figure}

After the fit, the spectator-isobar amplitude was analytically continued away from the real energy axis. The pole search was performed on the unphysical RSs reached by continuation across the relevant $K\bar K\pi$ and $(K\bar K^*)_+$ cuts, while using the corresponding analytic continuation of the isobar propagators described in Sec.~\ref{sec:2b}. The $\pi a_0$ cut is not encountered on these sheets: it would be reached by crossing the segment of the three-body cut between the $\eta\pi\pi$ and $K\bar K\pi$ thresholds, whereas the sheets studied here are accessed by crossing the cut above the $K\bar K\pi$ threshold. The analytic continuation of the spectator-isobar amplitude was implemented through contour deformation of the spectator momentum integration contour. The deformed contour was parametrized in the same form as the bSMC introduced in Eq.~\eqref{eq:bSMC}, but with a larger deformation into the lower half-plane ($w=0.2$~GeV, $V_0=-0.3$~GeV). A representative density plot of $|\mathcal T|$ on different RSs with the $K\bar K\pi$ and $(K\bar K^*)_+$ cuts is shown in Fig.~\ref{fig:pole_positions}.

The resulting pole positions are listed in Table~\ref{tab:res}. In particular, the pole associated with the $f_1(1285)$ has a small imaginary part. It can be compared with the resonance parameters extracted by BESIII from their full partial-wave analysis of the $J/\psi\to K_S^0K_S^0\pi^0$ (note that we only fit to the $1^{++}$ part of their extraction), which are $M=1280.2 \pm 0.6^{+1.2}_{-1.5}$~MeV and $\Gamma=28.2 \pm 1.1^{+5.5}_{-2.9}$~MeV~\cite{BESIII:2022chl}, as well as with the current averages in the Review of Particle Physics, $M=1281.8\pm0.5$~MeV and $\Gamma=23.0\pm1.0$~MeV~\cite{ParticleDataGroup:2026mpi}. The pole mass ($1277\pm2$~MeV) and width ($24\pm2$ MeV) obtained in the present analysis are in good agreement with these values, supporting the identification of this pole with the established $f_1(1285)$ resonance. Besides the $f_1(1285)$ and $f_1(1420)$ poles, the analytic continuation also reveals an extra pole on the same RS as the $f_1(1285)$, reached directly through the $K\bar K\pi$ cut. This pole lies substantially deeper in the complex plane than the other two poles. Consequently, it has little visible impact on the physical lineshape and is much less constrained by the fit. Its position shows a sizable dependence on the cutoff and on the two-body input, as it comes mainly from the $P$-wave $a_0\pi$ interaction (see below for further discussion).\footnote{To locate this pole when $\Lambda=0.73$~GeV, we set $V_0=-0.1$~GeV. A larger value of $|V_0|$ blocks access to this pole.}

\begin{figure}[tb]
    \centering
    \includegraphics[width=\textwidth]{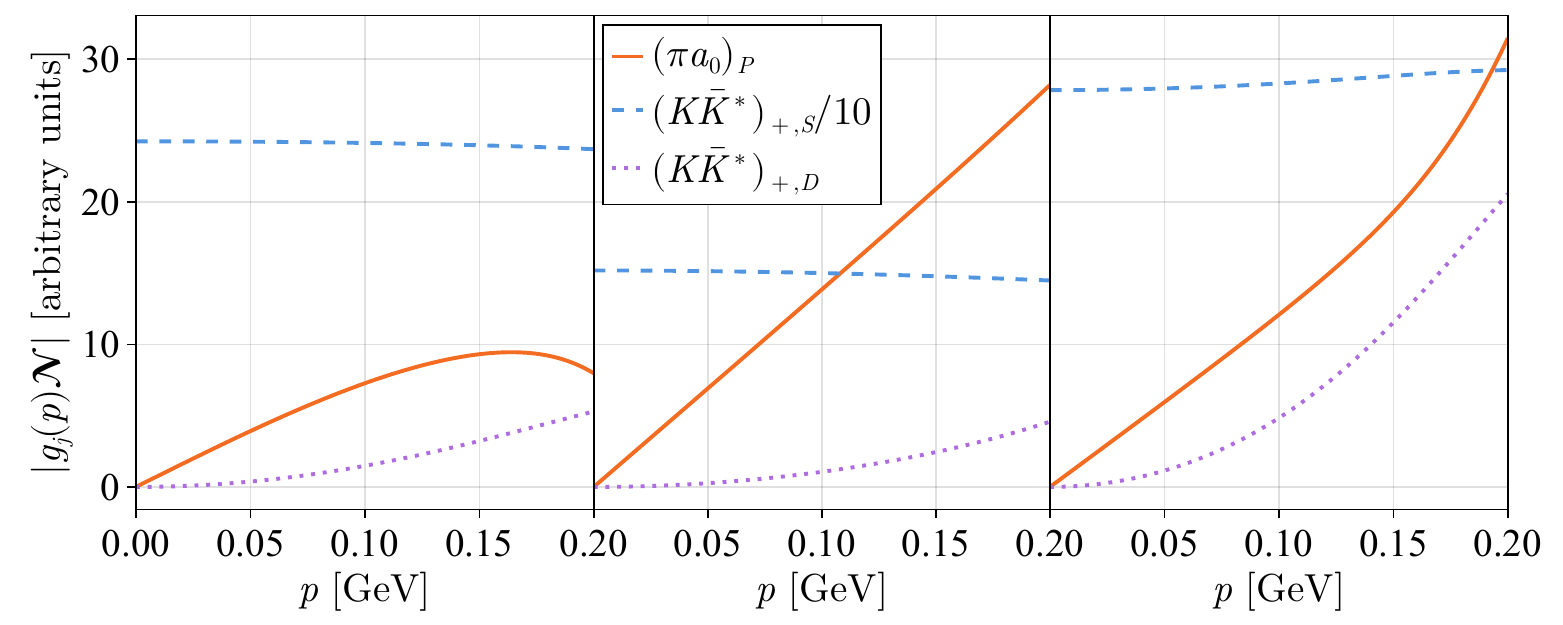}
    \caption{Momentum-dependent couplings of the three poles to three channels. The panels correspond to the $f_1(1285)$, $f_1(1420)$, and the extra pole, respectively. The $S$-wave $(K\bar K^*)_+$ couplings are scaled down by a factor of 10 for better display.}
    \label{fig:residues}
\end{figure}

To characterize the channel content of the poles, we also computed their couplings to the three spectator-isobar channels, $(\pi a_0)_P$, $(K\bar K^*)_{+,S}$, and $(K\bar K^*)_{+,D}$. Near a pole at $s=s_p$, the partial-wave amplitude factorizes as
\begin{equation}\label{eq:pole_factorization}
\mathcal T_{ji}^J(s,p^\prime,p)=\frac{g_j(p^\prime)g_i(p)}{s-s_p}+\text{regular~terms}\,.
\end{equation}
Substituting Eq.~\eqref{eq:pole_factorization} into the decay amplitude in Eq.~\eqref{eq:prod} gives
\begin{equation}\label{eq:prod_residue}
\lim_{s\to s_p}\frac{(s-s_p)\mathcal A_j^J\left(s,p^\prime\right)}
{\tau_j(\sigma_{p^\prime,j})}
=g_j(p^\prime)\,\mathcal N\,,\quad
\mathcal N=\sumint_i\frac{p^2\mathrm dp}{(2\pi)^32E_{p,i}}\,
g_i(p)\tau_i(\sigma_{p,i})\mathcal D_i^J(p)\,,
\end{equation}
up to terms regular at the pole. Thus the quantity $(s-s_p)\mathcal A_j^J/\tau_j$ is proportional to the pole residue in channel $j$, with a common $\mathcal D$-dependent normalization $\mathcal N$. This relation fixes the relative magnitudes of the couplings to the three channels. These couplings are displayed in Fig.~\ref{fig:residues}. One clearly observes that the couplings to the three channels exhibit the expected threshold behavior of the corresponding partial waves, i.e., $p^L$ with $L = 0, 1, 2$. Even though the contact potential $\mathcal C$ contains neither $D$-wave nor $S$-$D$ mixing interactions, the $D$-wave $(K\bar K^*)_+$ couplings to the poles remain non-zero. This originates from the one-particle-exchange $\mathcal B$, which does include $D$-wave and $S$-$D$ mixing interactions.

\begin{figure}[tb]
    \centering
    \includegraphics[width=0.75\textwidth]{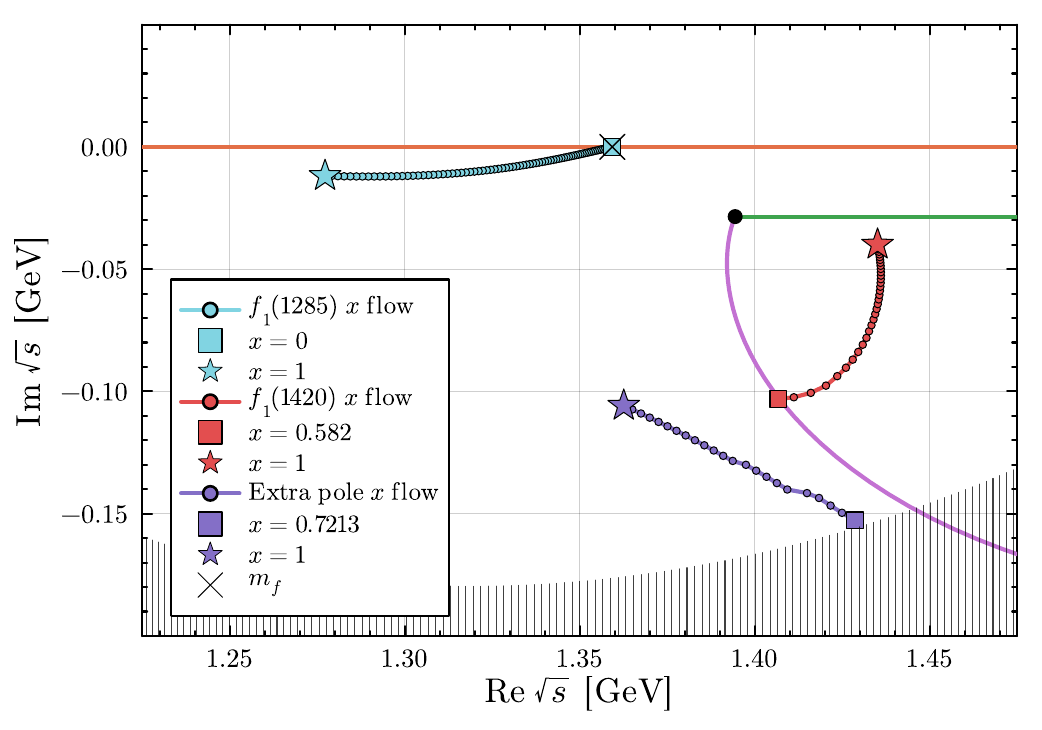}
    \caption{Trajectories of the three poles along the interpolation parameter $x$ defined in Eq.~\eqref{eq:xpara}. The orange and green lines represent the $K\bar{K}\pi$ and $(K\bar K^*)_+$ right-hand cuts, respectively. The black dot marks the $(K\bar K^*)_+$ branch point: $\sqrt{s}=m_K+\sqrt{\sigma_{K^*}}$. The magenta curve shows the mapping of the integration contour onto the $\sqrt{s}$ complex plane via Eq.~\eqref{eq:kin} with $\sigma=\sigma_{K^*}$. As $x$ decreases, the $f_1(1420)$ pole moves deeper into the complex plane and eventually approaches the boundary of the integration contour on the unphysical RS. The extra pole also moves toward a deeper region of the complex plane that is not accessible with the present integration contour (dense vertical hatching). Therefore neither dynamically generated trajectory can be followed all the way to $x=0$ in this plot, while the lightest pole trajectory remains unaffected by this limitation.}
    \label{fig:trajectory}
\end{figure}

To clarify the origin of the poles, we note that the contact potential $\mathcal C$ contains only a single bare state. It is therefore important to determine whether one of the poles originates from this bare state while the others are dynamically generated, or whether several poles emerge from a splitting of the same bare state after dressing effects are included. For this purpose, we introduce an interpolation parameter $x$ to smoothly connect the bare interaction ($x=0$) with the full physical amplitude ($x=1$):
\begin{equation}
\mathcal T^J_{ji}\left(s,p^\prime,p;x\right)=\mathcal V^J_{ji}\left(s,p^\prime,p\right)+x~\sumint_k\frac{q^2\mathrm{d}q}{(2\pi )^32E_{q,k}}\mathcal V^J_{jk}\left(s,p^\prime,q\right)\tau _k\left(\sigma _{q,k}\right)\mathcal T^J_{ki}\left(s,q,p;x\right).
\label{eq:xpara}
\end{equation}
Varying $x$ from $0$ to $1$ describes the dressing evolution of the poles; see the resulting pole trajectories in Fig.~\ref{fig:trajectory}. The lower pole associated with the $f_1(1285)$ continuously evolves from the bare-state pole as $x$ increases, indicating that the $f_1(1285)$ is predominantly from dressing the bare state in the present parametrization. In contrast, the higher pole, identified with the $f_1(1420)$, moves upward toward the physical region as $x$ increases; when $x$ is decreased, it moves deeper into the complex plane and eventually reaches the accessible boundary of the integration contour on the unphysical RS. Although its trajectory cannot be followed all the way to $x=0$ in the figure, it is expected that in this limit the pole would be located at infinity on the same RS. 
The extra pole behaves similarly: as $x$ decreases, it moves to a deeper region of the complex plane on the same sheet as the $f_1(1285)$, beyond the region accessible with the present contour deformation, so its trajectory also has to be terminated before reaching $x=0$. This behavior suggests that both the $f_1(1420)$ and the extra pole are dynamically generated, while only the $f_1(1285)$ is continuously connected to the bare state.

\begin{table}[tb]
    \caption{Pole content of single-channel limits of the reference-fit amplitude. ``$(K\bar K^*)_+$ only'' means that the $\pi a_0$ channel is fully decoupled ($c_{11}=c_{12}=h_1=0$ and kaon exchange switched off); ``$\pi a_0$ only'' means that the $(K\bar K^*)_+$ sector is switched off ($c_{12}=c_{22}=h_2=0$ and all one-particle exchange off; the diagonal $\pi a_0$ interaction receives no exchange contribution). ``Bare'' in the title line indicates whether the explicit bare-state term is kept. The poles are arranged in columns according to their physical origins, and ``--'' denotes that no corresponding pole is found in that limit. All poles are verified to be stable under variations of the contour-deformation depth.}
    \label{tab:origin}
    \center
    \renewcommand{\arraystretch}{1.25}
    \begin{tabular}{lcccc}
    \hline\hline
    \multirow{2}{*}{Limit} & \multirow{2}{*}{Bare} & \multicolumn{3}{c}{Poles [MeV]} \\
     & & $f_1(1285)$ & $f_1(1420)$ & extra pole \\
    \hline
    full amplitude                     & yes & $1277-\ii\,12$ & $1435-\ii\,40$  & $1363-\ii\,106$ \\
    $(K\bar K^*)_+$ only               & yes & $1303-\ii\,3$  & $1405-\ii\,21$  & --          \\
    $(K\bar K^*)_+$ only               & no  & --         & $1358-\ii\,16$  & --          \\
    $(K\bar K^*)_+$ only, contact only & no  & --         & $1381-\ii\,16$  & --          \\
    $\pi a_0$ only                     & yes & $1320-\ii\,40$ & --          & $1407-\ii\,85$  \\
    $\pi a_0$ only                     & no  & --         & --          & $1352-\ii\,125$ \\
    \hline\hline
    \end{tabular}
\end{table}

The global parameter $x$ in Eq.~\eqref{eq:xpara} dresses all channels simultaneously and therefore does not reveal the dynamics from which channel leads to the two dynamically generated poles. To disentangle this, we study single-channel limits of the amplitude and channel-resolved coupling scalings using the best-fit parameters from the reference fit in Table~\ref{tab:res}, with the resulting pole content collected in Table~\ref{tab:origin}.

First, we decouple the $\pi a_0$ channel entirely, switching off the $\pi a_0\leftrightarrow (K\bar K^*)_+$ contact $c_{12}$ and the kaon-exchange interaction, as well as setting $c_{11}=h_1=0$, so that only the $(K\bar K^*)_+$ sector remains. In this limit the spectrum consists of a narrow state at $(1303-\ii\,3)$~MeV and a broader one at $(1405-\ii\,21)$~MeV, which evolve into the physical $f_1(1285)$ and $f_1(1420)$ poles once the $\pi a_0$ coupling is restored. With the bare state also removed ($h_2=0$), the remaining interaction (the contact $c_{22}$ plus the pion-exchange term) still generates a pole at $\sqrt{s}=(1358-\ii\,16)$~MeV, below the $K\bar K^*$ threshold: a $K\bar K^*$ quasi-bound state whose width derives from the $K^*$ self-energy and is smaller than the $K^*$ width, which is similar to the case of the double-charm $T_{cc}(3875)$ obtained in the three-body framework (scheme II in Ref.~\cite{Du:2021zzh}). 
The pole also survives by keeping the contact interaction ($c_{22}$) alone, which yields a pole at $(1381-\ii\,16)$~MeV. 
Thus, we conclude that the $f_1(1420)$ emerges predominantly as a $K\bar K^*$ dynamically generated state, supporting its hadronic-molecule interpretation~\cite{Longacre:1990uc, Tornqvist:1993ng}, with an admixture of the nearby bare state, while the $f_1(1285)$ is the corresponding dressed bare state, consistent with the trajectories in Fig.~\ref{fig:trajectory}. 

The extra pole has a different origin. It is absent from the $(K\bar K^*)_+$ sector in all of the above limits. Instead, keeping only the $\pi a_0$ channel with its $P$-wave contact $c_{11}$ (in this limit there is no one-particle exchange, and the bare state is removed) yields a single broad pole at $(1352-\ii\,125)$~MeV, close to the extra pole of the full amplitude at $(1363-\ii\,106)$~MeV. This pole is therefore generated by the $\pi a_0$ $P$-wave contact interaction alone and is not a shadow pole of the $f_1(1420)$ arising from the $K\bar K^*$ dynamics. This origin also explains its pronounced sensitivity to the cutoff and to the two-body input noted above: a short-range $P$-wave attraction is strongly scheme dependent, and we find that indeed scaling down $c_{11}$ by only ${\sim}30\%$ already moves this pole by $\mathcal{O}(40)$~MeV and expels it from the region accessible to our contour deformation, while the $f_1(1285)$ and $f_1(1420)$ poles shift by only a few MeV.
In contrast, scaling down $c_{22}$ alone by a factor from 1.0 to 0.1 broadens the $f_1(1420)$ width from $\Gamma\simeq 80$ to $\simeq 175$~MeV, while the extra pole barely responds, moving by less than $25$~MeV over the full scaling range.

\begin{figure}[tb]
    \centering
    \includegraphics[width=0.75\textwidth]{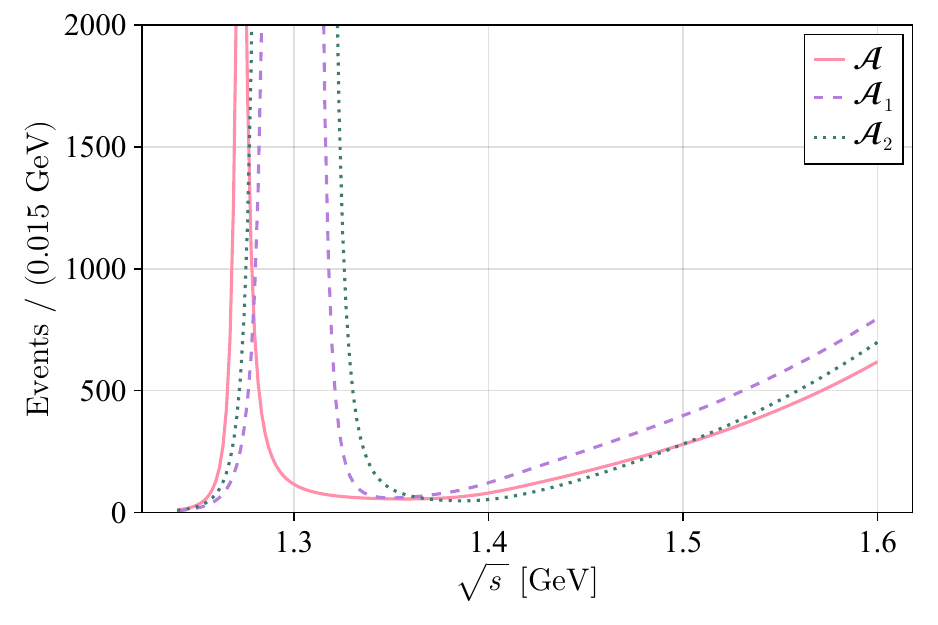}
    \caption{Auxiliary lineshape obtained by setting $c_{11}=c_{12}=c_{22}=0$ and $m_f=1.3$~GeV. Here $\mathcal A$ denotes the decay amplitude with the full infinite-order spectator-isobar resummation, while $\mathcal A_N=\tau\sum_{n=0}^{N}(\mathcal V\tau)^n\mathcal D$ denotes the amplitude truncated after $N$ rescattering insertions. In this setup the sequence from $\mathcal A_1$ to $\mathcal A_2$ and then to $\mathcal A$ is already convergent, indicating that the poor order-by-order convergence near $1.43$~GeV in Fig.~\ref{fig:fit_lineshape} is associated with the nearby dynamically generated $f_1(1420)$ pole rather than with the triangle singularity alone.}
    \label{fig:lineshape_no_pole}
\end{figure}

\begin{figure}[tb]
    \centering
    \includegraphics[width=0.75\textwidth]{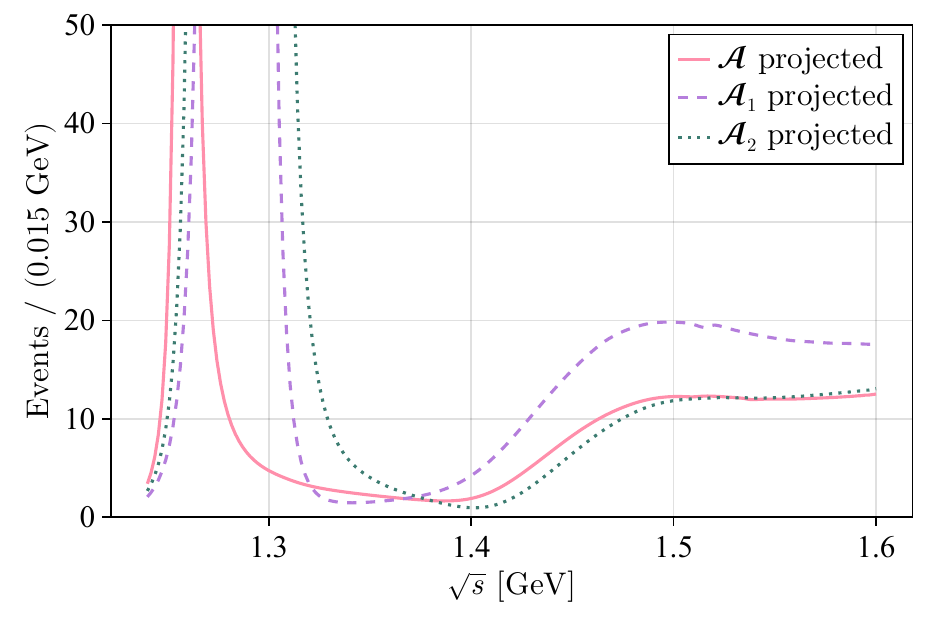}
    \caption{Projected auxiliary lineshape obtained with the same parameters as in Fig.~\ref{fig:lineshape_no_pole}, but with the $K\bar K$ invariant mass integrated only from threshold to 10~MeV above threshold. The labels $\mathcal A$ and $\mathcal A_N$ have the same meaning as in Fig.~\ref{fig:lineshape_no_pole}; ``projected'' indicates that the three-body phase space is restricted to the near-threshold $K\bar K$ region, which mimics the triangle-singularity contribution to $f_1(1285)\to\pi a_0$. This projection makes the triangle singularity visible as a rapid rise starting around $1.40$~GeV, while the convergence from $\mathcal A_1$ to $\mathcal A_2$ and then to $\mathcal A$ remains good.}
    \label{fig:lineshape_no_pole_proj}
\end{figure}

As discussed above, the rescattering series does not converge in the $1.43$~GeV region, in contrast to the rapidly converging case of Ref.~\cite{Sakthivasan:2024uwd}. This difference can be traced to the pole content of the amplitude: the present fit contains the dynamically generated $f_1(1420)$ pole close to the $K\bar K^*$ threshold, whereas the amplitude of Ref.~\cite{Sakthivasan:2024uwd} does not. The structure around the $K\bar K^*$ threshold therefore cannot be treated perturbatively and requires the full infinite-order resummation. This interpretation is supported by the auxiliary calculations in Figs.~\ref{fig:lineshape_no_pole} and~\ref{fig:lineshape_no_pole_proj}. After removing the nearby dynamically generated pole by setting $c_{11}=c_{12}=c_{22}=0$ and shifting the bare mass to $m_f=1.3$~GeV (chosen only to push the bare state slightly toward lower energies so that it does not obscure the lineshape in the triangle-singularity region), the sequence from $\mathcal A_1$ to $\mathcal A_2$ and then to the fully resummed $\mathcal A$ becomes convergent. The fully integrated lineshape in Fig.~\ref{fig:lineshape_no_pole}, however, does not exhibit a sharp triangle-singularity signal, because the three-body phase-space integration smears it out. Once the $K\bar K$ invariant mass is projected onto the near-threshold region by integrating only the $K\bar K$ invariant mass from the threshold to 10~MeV above threshold, the triangle-singularity contribution becomes evident as a bump above 1.4~GeV, as shown in Fig.~\ref{fig:lineshape_no_pole_proj}. This is because the triangle-singularity effect is sensitive to kinematical variables and is particularly strong when the $K\bar K$ invariant mass is near the threshold (for a detailed discussion, we refer to Section~4.2 in Ref.~\cite{Guo:2019twa}).

\begin{figure}[tb]
    \centering
    \includegraphics[width=0.75\textwidth]{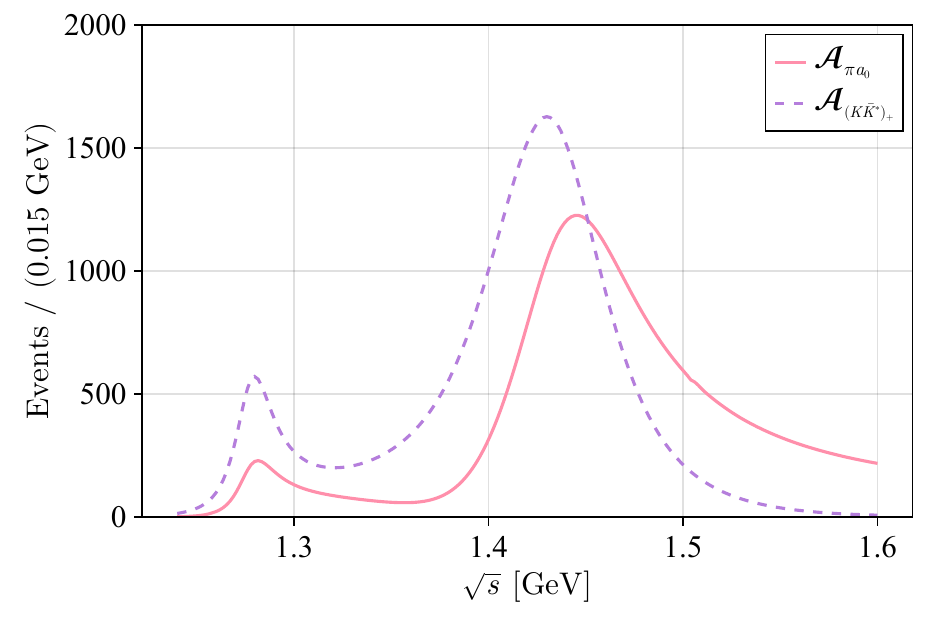}
    \caption{Lineshapes corresponding to the $\pi a_0$ and $(K\bar K^*)_+$ terms in Eq.~(\ref{eq:finalA}). The physical $K\bar K\pi$ lineshape is obtained from their coherent sum. The individual curves are shown only for illustration and do not represent separately observable contributions.}
    \label{fig:lineshape_last_only}
\end{figure}

Finally, unlike in a conventional isobar model, the $\pi a_0$ and $(K\bar K^*)_+$ terms in Eq.~\eqref{eq:finalA} are not independently parametrized components. They are two terms of a single coupled-channel decay amplitude and are mixed by the one-particle-exchange and short-range interactions through the all-order rescattering. Their coherent sum gives the physical $K\bar K\pi$ amplitude, while the individual terms are not separately observable. Nevertheless, we show in Fig.~\ref{fig:lineshape_last_only} the lineshapes obtained from the $\pi a_0$ and $(K\bar K^*)_+$ terms in Eq.~\eqref{eq:finalA} separately, before their coherent summation. These curves should not be interpreted as independent physical contributions in the sense of a conventional isobar model, but merely illustrate how the final coupled-channel amplitude is decomposed into its two formal terms.

\section{Summary}

We have studied the $I^G(J^{PC})=0^+(1^{++})$ $K\bar K\pi$ system in a relativistic three-body unitary framework. The explicit channels retained in the calculation are the $P$-wave $\pi a_0$ channel as well as the $S$- and $D$-wave $(K\bar K^*)_+$ channels. The one-particle-exchange interaction, fixed by three-body unitarity, generates the long-range part of the amplitude and automatically contains the triangle-singularity mechanism associated with the $K\bar K \bar K^*$ loop. The remaining short-range three-body interaction was fitted to the $0^+(1^{++})$ component of the BESIII $K^0_SK^0_S\pi^0$ invariant-mass distribution.

After fitting the physical lineshape, we analytically continued the amplitude to the relevant unphysical RSs using deformed integration contours. Two poles were found robustly under variations of the cutoff and of the $K\bar K$ scalar input. Combining the statistical uncertainty of the reference fit with the spread among the three schemes, the positions are
\begin{align}
\sqrt{s_{f_1(1285)}}&=
\left(1277\pm2\pm1\right)
-\ii\left(12\pm1\pm0\right)\text{MeV}\,,\notag\\
\sqrt{s_{f_1(1420)}}&=
\left(1435\pm2\pm7\right)
-\ii\left(40\pm2\pm1\right)\text{MeV}\,,
\end{align}
where the first uncertainties are statistical, while the second uncertainties estimate the systematic effects from the two model-dependence checks. 
The central values of the corresponding reference-fit widths are about $24$ MeV and $80$ MeV, respectively. 
The analytic continuation also reveals an additional pole on the same RS as the $f_1(1285)$. Since this pole lies deeper in the complex plane and exhibits sizable cutoff and two-body-input dependence, we regard it to be more model-dependent than the above two poles. The pole-trajectory analysis indicates different origins for the poles. The $f_1(1285)$ pole evolves continuously from the bare state introduced in the potential. The $f_1(1420)$ pole and the extra pole, by contrast, are dynamically generated by resumming the coupled-channel amplitude via the three-body unitary framework.
The single-channel limits collected in Table~\ref{tab:origin} further resolve their channel origins. The $(K\bar K^*)_+$ interaction alone generates a quasi-bound state below the $K\bar K^*$ threshold; mixing with the nearby bare state then yields the two levels that evolve into the physical $f_1(1285)$ and $f_1(1420)$ poles. The $f_1(1420)$ can therefore be regarded as predominantly an $S$-wave $K\bar K^*$ molecular state with a bare-state admixture. 
The extra pole, in contrast, is generated by the $P$-wave $\pi a_0$ contact interaction alone and is not a shadow pole of the $f_1(1420)$ from the $K\bar K^*$ dynamics. The strong scheme dependence inherent to such a short-range $P$-wave attraction naturally explains its sizable cutoff and two-body-input dependence. The presence of the $f_1(1420)$ pole implies that the peak around 1.43~GeV of the $K_S^0K_S^0\pi^0$ invariant-mass distribution of the BESIII data~\cite{BESIII:2022chl} is not merely a triangle-singularity enhancement, although the triangle singularity plays a role in the region around $1.40$--$1.42$~GeV.

It is worth stressing that an early analysis of the $f_1(1420)$ already employed a three-body picture and considered the coupled $\pi a_0$ and $K\bar K^*$ dynamics~\cite{Longacre:1990uc}. The present work follows the same physical motivation, but implements it with a modern infinite-volume unitary amplitude, controlled analytic continuation, and an explicit pole search. In this sense, the triangle singularity and the coupled-channel dynamics are not treated as alternatives to pole physics; instead, they are incorporated into the same amplitude, allowing the pole content of the region of the $f_1(1285)$ and $f_1(1420)$ to be determined consistently.

\begin{acknowledgments}

We are grateful to the BESIII Collaboration for providing the lineshape data. The authors also thank Maxim Mai, Daniel Sadasivan, and Yi-Ling Song for useful discussions.
This work is supported in part by the National Key R\&D Program of China under Grant No. 2023YFA1606703; by the National Natural Science Foundation of China (NSFC) under Grants No. 12125507, No. 12361141819, No. 12405106, and No. 12447101; and by the Chinese Academy of Sciences (CAS) under Grant No.~YSBR-101.
U.-G.M. also acknowledges the support from the Deutsche Forschungsgemeinschaft (DFG) under Germany's Excellence Strategy -- EXC 3107 -- Project-ID~533766364, and from the CAS President's International Fellowship Initiative (PIFI) under Grant No.~2025PD0022.

\end{acknowledgments}

\appendix

\section{Contour deformation method}\label{app:CD}

This appendix is a technical supplement to the lineshape construction rather than part of the main numerical workflow, in which the fits use the AAA procedure. Here we explain how the same real-axis decay amplitudes $\mathcal A^J_j(s,p^\prime)$ can be obtained rigorously by contour deformation.

The discussion again focuses directly on the decay amplitude $\mathcal A^J_j(s,p^\prime)$ for $p^\prime \in \mathbb{R}$. This is the primary quantity for calculating the physical lineshape, and its direct determination allows us to bypass the full $\mathcal{T}^J_{ji}(s,p^\prime,p)$, thereby avoiding the unnecessary complications associated with the incoming spectator momentum $p$. Let us start with
\begin{equation}\label{eq:prod4}
\mathcal A^J_{j}\left(s,p^\prime\right)=\tau_j\left(\sigma_{p^\prime,j}\right)\left(\mathcal D^J_{j}\left(p^\prime\right)+\sumint_{i,p\in\text{SMC}_i\left(s,p^\prime\right)}\frac{p^2\mathrm{d}p}{(2\pi )^32E_{p,i}}\mathcal V_{ji}^J\left(s,p^\prime,p\right)\mathcal A^J_{i}\left(s,p\right)\right).
\end{equation}
Here, the integration contour for $p$, namely the spectator momentum contour (SMC) $\text{SMC}_i(s,p^\prime)$, is chosen for a fixed channel $i$ and a fixed real $p^\prime$. To evaluate this expression, one must first determine the values of $\mathcal A^J_i(s,p)$ for $p$ residing on $\text{SMC}_i(s,p^\prime)$. These are obtained via
\begin{equation}\label{eq:prod5}
\mathcal A^J_{i}\left(s,p\right)=\sumint_{k,\,q\in\text{AIC}(p)}\mathrm d q\left(1-\tau\mathcal V^J \right)^{-1}_{ik}\left(s,p,q\right)\tau_k\left(\sigma_{q,k}\right)\mathcal D^J_{k}(q)\,.
\end{equation}
Consequently, two distinct contour problems must be addressed:
\begin{enumerate}
\item $\text{SMC}_i(s,p^\prime)$ employed in Eq.~\eqref{eq:prod4} for a fixed real $p^\prime$;
\item the auxiliary inversion contour (AIC) $\text{AIC}(p)$ employed in Eq.~\eqref{eq:prod5} to determine $\mathcal A^J_i(s,p)$ for the required set of target points $p$, with the contour chosen such that $p\in \text{AIC}(p)$. In general, different target points $p$ may require different $\text{AIC}(p)$. In special cases, however, a single contour can pass through all points on a given $\text{SMC}_i(s,p^\prime)$, in which case one may identify $\text{AIC}(p)=\text{SMC}_i(s,p^\prime)$ for all such $p$.
\end{enumerate}

These integration contours must avoid all singularities in the complex momentum plane. The most important and non-trivial among them are the three-body cuts in momentum space, which are simply another manifestation of the same right-hand $K\bar K\pi$ cut appearing in the complex $\sqrt{s}$ plane, expressed in a different variable. In the following, we shall refer to them simply as the three-body cuts.

The three-body cuts originate from the one-particle-exchange term $\mathcal B_{ji}(s,p^\prime,p)$. For fixed $s$ and $p^\prime$, the corresponding singularities in the complex $p$ plane are determined by
\begin{equation}
\sqrt{s}-E_{\boldsymbol{p},i}-E_{\boldsymbol{p}^\prime,j}-E_{\boldsymbol{p}+\boldsymbol{p}^\prime,\text{ex}}=0\,,
\end{equation}
whose solutions read as~\cite{Fix:2001cz, Zhang:2024dth, Feng:2024wyg}
\begin{align}\label{eq:lhc}
p_{\pm}\left(s,p^\prime,z\right)&=\frac{-2\alpha p^\prime z\pm \sqrt{4\alpha ^2p^{\prime2}z^2+\left(\beta ^2-4p^{\prime2}z^2\right)\left(\alpha ^2-\beta ^2m_{i}^{2}\right)}}{\beta ^2-4p^{\prime2}z^2},\notag\\
\alpha&=s+m_i^2+m_j^2-m_\text{ex}^2-2\sqrt{s}E_{p^\prime,j}\,,\notag\\
\beta&=2\left(\sqrt{s}-E_{p^\prime,j}\right),
\end{align}
with $z=\cos\theta\in[-1,1]$. As $z$ varies, these solutions generate the three-body cuts. The deformation of the integration contour in response to such moving singularities, and the contour pinching that produces a triangle singularity, are discussed pedagogically in Ref.~\cite{Aitchison:2015jxa}.

In the present discussion, the one-particle exchange $\mathcal{B}$ appears in two places: $\mathcal{V}_{ji}^J(s,p^\prime,p)$ in Eq.~\eqref{eq:prod4} and $(1 - \tau \mathcal{V}^J)^{-1}_{ik}(s,p,q)$ in Eq.~\eqref{eq:prod5}. The integration contours must be chosen carefully to avoid all singularities arising from these contributions. This imposes the following two conditions:
\begin{enumerate}
\item for Eq.~\eqref{eq:prod4}, $\text{SMC}_i(s,p^\prime)$ must avoid the cuts defined by $p_\pm(s,p^\prime,z\in[-1,1])$;
\item for Eq.~\eqref{eq:prod5}, $\text{AIC}(p)$ must satisfy a stronger condition: if the inversion is performed along a contour $\text{AIC}(p)$ in the complex $q$ plane, then for every $q\in\text{AIC}(p)$ the non-analytic regions generated by $p_\pm(s,q,z\in[-1,1])$ must again be avoided by the same contour. This leads to the practical rule that the AIC should be deformed exclusively into either the upper or lower half-plane. It cannot remain entirely on the real axis, nor should it typically traverse both half-planes simultaneously (see, e.g., the contour shape in Fig.~\ref{fig:lhc2} below, which will be discussed in further detail later).
\end{enumerate}

\begin{figure}[tb]
    \centering
    \includegraphics[width=\linewidth]{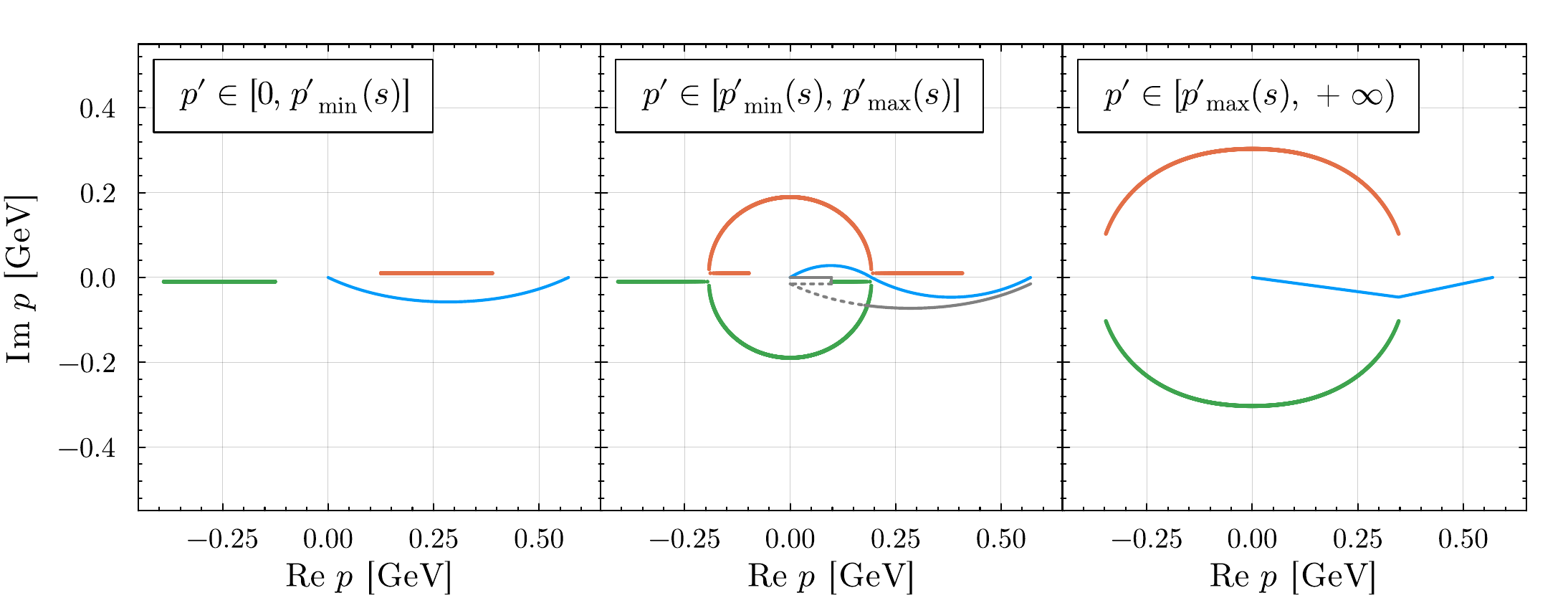}
    \caption{Three-body cuts in the complex $p$ plane and possible choice of the integration contour SMC. The three panels correspond to three different regions of $p^\prime$. The orange and green curves represent $p_\pm(s,p^\prime,z\in[-1,1])$ in Eq.~\eqref{eq:lhc}, respectively. The blue curve denotes a possible choice of integration contour in each case. The gray curve indicates an alternative contour when $p^\prime \in [p^\prime_{\min}(s), p^\prime_{\max}(s)]$. }
    \label{fig:lhc}
\end{figure}

First, let us discuss condition~1. The real $p^\prime$ axis can be divided into three regions:
$p^\prime \in [0, p^\prime_{\min}(s)]$, $[p^\prime_{\min}(s), p^\prime_{\max}(s)]$, and $[p^\prime_{\max}(s), +\infty)$.
The boundaries are given by~\cite{Fix:2001cz, Zhang:2024dth, Feng:2024wyg}
\begin{equation}
p^\prime_{\min}(s) = p_{\mathrm{cm}}\left(\sqrt{s} - m_i;\, m_{\text{ex}}, m_j\right), \quad
p^\prime_{\max}(s) = p_{\mathrm{cm}}\left(\sqrt s;\, m_i + m_{\text{ex}}, m_j\right).
\end{equation}
For non-relativistic expressions, see, e.g., Refs.~\cite{Hetherington:1965zza, Aaron:1966zz, Brayshaw:1968yia, Cahill:1971ddy, Schmid:1974}. Note that for $p^\prime \in [p^\prime_{\max}(s), +\infty)$, the kinematics implies $\sqrt{\sigma_{p^\prime,j}} \leq m_i + m_{\text{ex}}$. Therefore, this region is not encountered in the phase-space integration of the decay amplitude. Nevertheless, we include it here for completeness. Within these three regions of $p^\prime$, the three-body cuts appearing in condition~1 exhibit three distinct topologies. Since for fixed $s$ and $p^\prime$ the cuts are fully determined, one can always construct a class of contours that satisfy condition~1; see Fig.~\ref{fig:lhc}. Once $\text{SMC}_i(s,p^\prime)$ is specified for given $s$ and $p^\prime$, one must precompute $\mathcal A^J_i(s,p)$ for $p$ residing on this contour, at which point the satisfaction of condition~2 becomes critical. We now describe this procedure in detail.

\subsection{Three regions}

\begin{figure}[tb]
    \centering
    \includegraphics[width=0.8\linewidth]{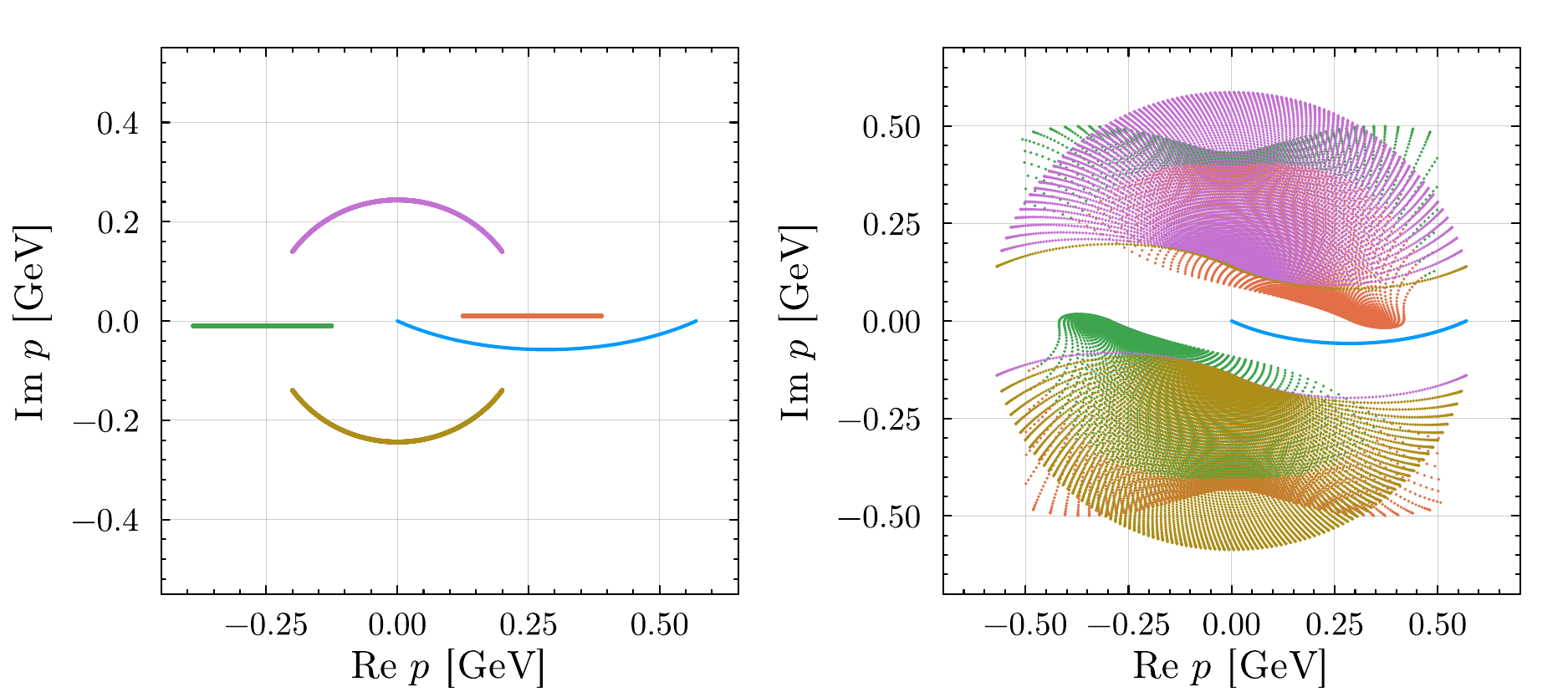}
    \caption{For a fixed value of $p^\prime\in [0, p^\prime_{\min}(s)]$, the singularity structure of the one-particle-exchange term (left), and the non-analytic regions generated by all momenta along the integration contour (right). The orange and green curves represent $p_\pm$ in Eq.~\eqref{eq:lhc}, while the magenta and brown curves denote $p_{\pm,\mathrm{ex}}$ in Eq.~\eqref{eq:lhcex}. The blue curve corresponds to the integration contour SMC. In both panels, the singularities remain separated from the integration contour.}
    \label{fig:lhc1}
\end{figure}

For $p^\prime \in [0, p^\prime_{\min}(s)]$, the two three-body cuts lie on the real axis. In this case, $\text{SMC}_i(s,p^\prime)$ can be chosen as a fixed deformation into the lower half-plane. The same contour is also admissible for the inversion in Eq.~\eqref{eq:prod5}: being one-sided, it avoids the non-analytic regions generated by itself. We therefore use the bSMC defined in Eq.~\eqref{eq:bSMC}, shown in Fig.~\ref{fig:lhc1}. In this first region, the bSMC serves as both the $\text{SMC}_i(s,p^\prime)$ in Eq.~\eqref{eq:prod4} and the AIC in Eq.~\eqref{eq:prod5}. Consequently, a single application of Eq.~\eqref{eq:prod5} suffices to determine all $\mathcal A^J_i(s,p)$ required as input for Eq.~\eqref{eq:prod4}. These precomputed values can be stored and reused for all $p^\prime \in [0, p^\prime_{\min}(s)]$, such that the calculation of $\mathcal A^J_j(s,p^\prime)$ reduces to a simple integration over these fixed quantities.

For $p^\prime \in [p^\prime_{\min}(s), p^\prime_{\max}(s)]$, the three-body cuts merge into a circular structure that nearly encircles the origin. In this case, which we call the critical region, $\text{SMC}_i(s,p^\prime)$ must be carefully deformed in the complex plane so as to connect $0$ and $\Lambda$ while avoiding the cuts. Figure~\ref{fig:lhc} shows two possible choices of the integration contour: the blue contour is used in Refs.~\cite{Cahill:1971ddy, Zhang:2024dth, Fix:2019txp}, while the gray contour is adopted in Refs.~\cite{Hetherington:1965zza, Aaron:1966zz, Schmid:1974, Fix:2001cz, Pang:2023jri}. We will discuss both of them in the next subsection.

\begin{figure}[tb]
    \centering
    \includegraphics[width=0.8\linewidth]{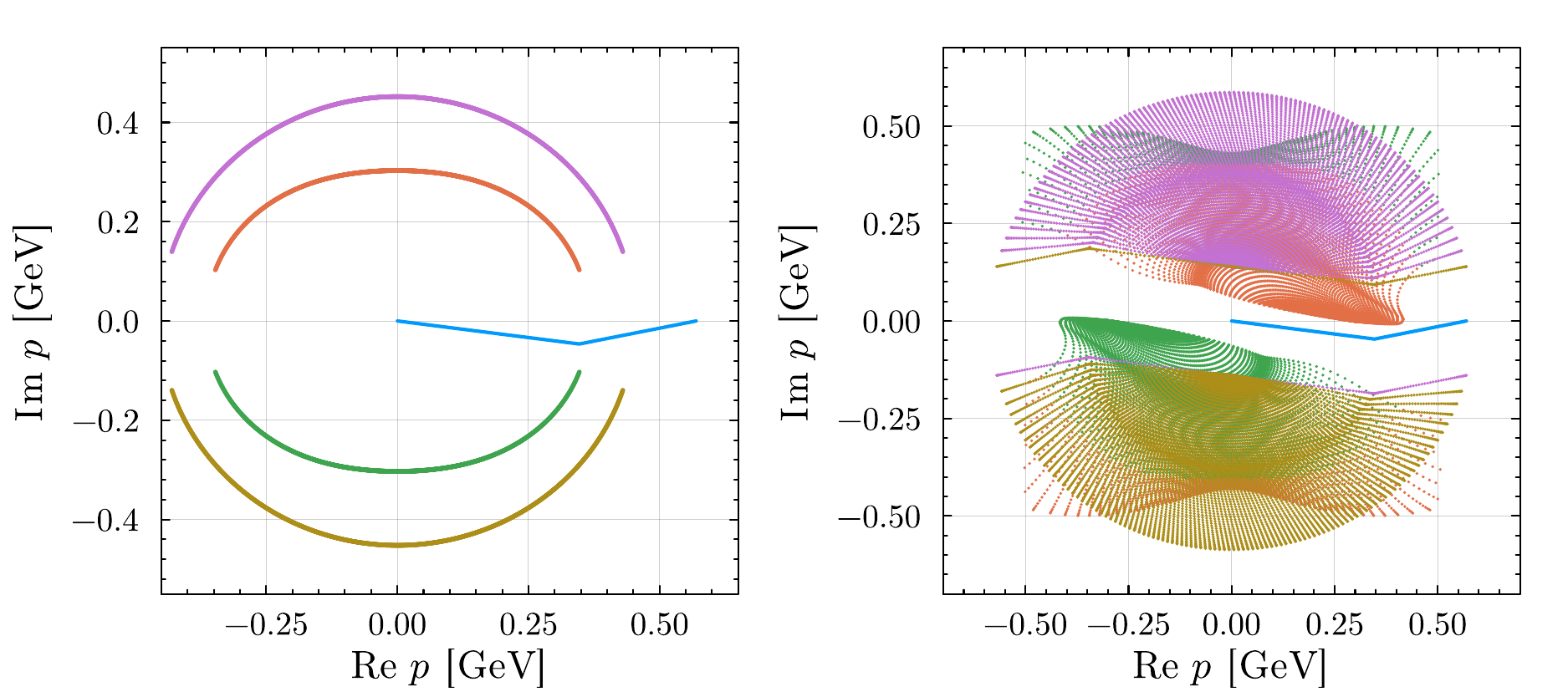}
    \caption{Same as Fig.~\ref{fig:lhc1}, but for $p^\prime \in [p^\prime_{\max}(s), +\infty)$.}
    \label{fig:lhc3}
\end{figure}

Finally, for $p^\prime \in [p^\prime_{\max}(s), +\infty)$, the circular cuts open up and the two cuts move completely away from the real axis. In this region, $\text{SMC}_i(s,p^\prime)$ in Eq.~\eqref{eq:prod4} may be chosen as a piecewise path running from $(0,0)$ to $\left(\mathrm{Re}\,p_{-}(s,p^\prime,-1),\, \mathrm{Im}\, p_{-}(s,p^\prime,-1)/2\right)$ and then to $(\Lambda,0)$; see Fig.~\ref{fig:lhc3}. Since this contour stays entirely in the lower half-plane apart from the endpoints, it can also be used as a one-sided AIC for Eq.~\eqref{eq:prod5}. Hence, as in the first region, $\mathcal A^J_j(s,p^\prime)$ can be obtained directly once the appropriate $\text{SMC}_i(s,p^\prime)$ is chosen.

\subsection{The critical region}

We now turn to the detailed discussion of the region $p^\prime \in [p^\prime_{\min}(s), p^\prime_{\max}(s)]$. To fully elucidate the underlying issues, we consider three different approaches. Two of them will eventually fail; however, these failures are instructive, as they shed light on the origin of the complexity in the three-body problem.

\begin{figure}[tb]
    \centering
    \includegraphics[width=0.8\linewidth]{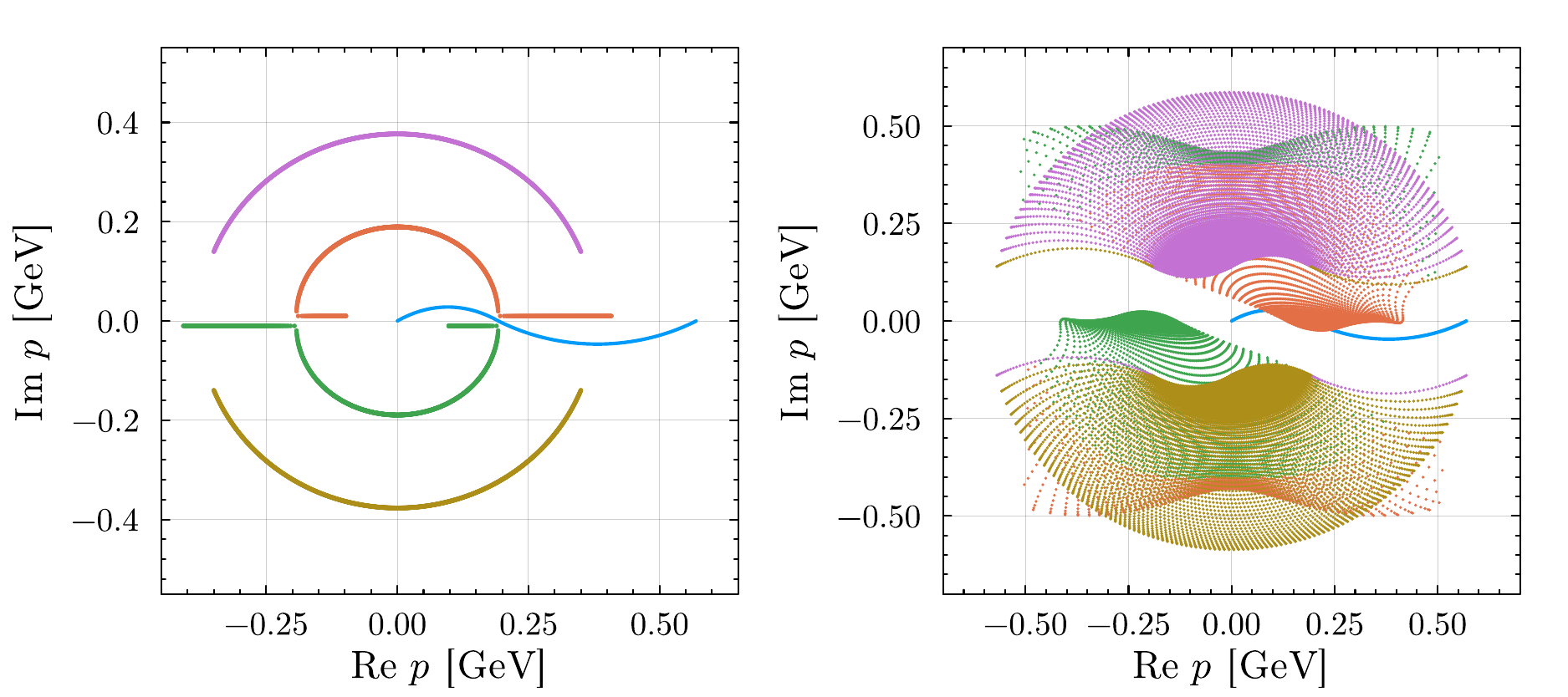}
    \caption{Same as Fig.~\ref{fig:lhc1}, but for $p^\prime \in [p^\prime_{\min}(s), p^\prime_{\max}(s)]$. In this case, the integration contour intersects the non-analytic regions generated by itself.}
    \label{fig:lhc2}
\end{figure}

\paragraph*{Method~1 (fails).} In this region, there exists a class of contours that satisfy condition~1. The contour must pass through the unique gap in the circular cuts, located at
\begin{equation}
p_{\text{gap}}\left(s,p^\prime\right) = \frac{m_i \sqrt{\beta^2 m_i^2 - \alpha^2}}{\alpha},
\end{equation}
with $\alpha$ and $\beta$ defined in Eq.~\eqref{eq:lhc},
shown in Fig.~\ref{fig:lhc2}. This contour is suitable for $\text{SMC}_i(s,p^\prime)$ in Eq.~\eqref{eq:prod4}, but it is not suitable as a global AIC for Eq.~\eqref{eq:prod5}: because it extends into both half-planes, there are values of $s$ and $p^\prime$ for which it intersects the non-analytic regions generated by itself; see Fig.~\ref{fig:lhc2}. In other words, a direct use of $\text{SMC}_{i}(s,p^\prime)$ in the inversion would fail.

\paragraph*{Method~2 (works).} The resolution is to separate the two integration contours. Equation~\eqref{eq:prod4} is evaluated on $\text{SMC}_{i}(s,p^\prime)$, but for each individual point $p\in\text{SMC}_{i}(s,p^\prime)$ the quantity $\mathcal A^J_i(s,p)$ is obtained from Eq.~\eqref{eq:prod5} using a specifically constructed $\text{AIC}(p)$. For any fixed target point $p$, one can choose an $\text{AIC}(p)$ passing through $p$ and lying entirely in the upper (lower) half-plane when $\mathrm{Im}\,p>0$ ($<0$). By the rule stated above, such one-sided contours satisfy condition~2. While this implies that a different inversion contour is required for each $p$, it ensures the analyticity of the kernel.

A minor caveat concerns the point $p_{\mathrm{gap}}(s,p^\prime)\in \mathrm{SMC}_{i}(s,p^\prime)$, which lies on the real axis. The value $\mathcal A_i^J(s,p_{\mathrm{gap}}(s,p^\prime))$ cannot be obtained by an AIC of the same one-sided type. This does not constitute a practical obstruction, however, since it is only an isolated point. In numerical implementation, one may choose the quadrature nodes such that this point is not sampled.

\paragraph*{Method~3 (fails).} We now turn to a third approach. Its underlying idea is similar to that of Method~2, but we now consider an alternative integration contour, i.e., the gray curve shown in Fig.~\ref{fig:lhc}, which is more commonly used in the literature.

Specifically, the integration proceeds from $0$ to $p_{-}(s,p^\prime,-1)$, where the contour crosses the branch point and moves onto the second RS defined by the three-body cut. It then returns to $0$, and subsequently runs from $0$ to $\Lambda$ along a contour deformed into the lower half-plane, which can be conveniently chosen as the bSMC. Along this second segment, the contour crosses the three-body cut once more, thereby returning to the first RS. When the contour enters the second RS of $\mathcal{B}_{ji}^J(s,p^\prime,p)$, one must use its value on that RS. This, however, does not affect $\mathcal A_{i}^J(s,p)$, since the RS structure under consideration is defined solely by the three-body cuts of $\mathcal{B}_{ji}^J(s,p^\prime,p)$.

To implement this integration contour, one needs prior knowledge of $\mathcal A_{i}^J(s,p)$ for $p$ within $[0, p_{-}(s,p^\prime,-1)]$, as well as for $p\in\text{bSMC}$. The values of $\mathcal A_{i}^J(s,p)$ with $p$ on the bSMC have already been obtained as in the first region. The remaining question is whether $\mathcal A_{i}^J(s,p)$ can be obtained for $p \in [0, p_{-}(s,p^\prime,-1)]$. From the previous discussion, we know that $\mathcal A_{i}^J(s,p)$ is directly accessible when $p \in [0, p^\prime_{\min}(s)]$; thus, Method~3 is applicable only if, for $p^\prime \in [p^\prime_{\min}(s), p^\prime_{\max}(s)]$, the condition
\begin{equation}
p_{-}\left(s,p^\prime,-1\right) < p^\prime_{\min}(s)
\end{equation}
is satisfied. Since $p_{-}(s, p^\prime, -1)$ attains its maximum value at $p^\prime = p^\prime_{\max}(s)$ within the interval $[p^\prime_{\min}(s), p^\prime_{\max}(s)]$, the above requirement is equivalent to
\begin{equation}
p_{-}\left(s,p^\prime_{\max}(s),-1\right) < p^\prime_{\min}(s)\,,
\end{equation}
which we now analyze.

First, one finds
\begin{equation}
p_{-}\left(s,p^\prime_{\max}(s),-1\right)=\frac{m_i}{m_i+m_{\text{ex}}}p^\prime_{\max}(s)\,.
\end{equation}
Then we introduce the ratio
\begin{align}
R(s)&=\left(\frac{p_{-}\left(s,p^\prime_{\max}(s),-1\right)}{p^\prime_{\min}(s)}\right)^2\\
&=\frac{m_i^2\left(\sqrt s-m_i\right)^2\left(\sqrt s+m_i+m_j+m_{\text{ex}}\right)\left(\sqrt s+m_i-m_j+m_{\text{ex}}\right)}{\left(m_i+m_{\text{ex}}\right)^2s\left(\sqrt s-m_i+m_j+m_{\text{ex}}\right)\left(\sqrt s-m_i-m_j+m_{\text{ex}}\right)}.
\end{align}
For equal masses $m_i=m_j = m_{\text{ex}}\equiv m$ (e.g., $3\pi$), one has
\begin{equation}
R(s)=\frac{\left(\sqrt s-m\right)\left(\sqrt s+3m\right)}{4s}<1\,,
\end{equation}
so the method is always admissible. For unequal masses, however, the condition must be checked case by case: for given particle masses and a specified range of $s$, one needs to check whether the condition $R(s)<1$ is satisfied within that range.

Here, we briefly examine the condition $R(s_{\text{thr}})<1$ at the threshold $\sqrt{s_{\text{thr}}}\equiv m_i+m_j+m_{\text{ex}}$, since this condition must be fulfilled in order to use the above integration contour:
\begin{equation}\label{eq:cond}
m_{\text{ex}}^3+\left(2m_i+m_j\right)m_{\text{ex}}^2+m_im_jm_{\text{ex}}-m_i^2m_j>0\,.
\end{equation}

Let us now see how this condition simplifies in the case where two of the three particles have the same mass. Denoting the distinct mass by $m_a$ and the two identical masses by $m_b$, one may consider the three possible assignments obtained by permuting $m_a$ among incoming $m_i$, outgoing $m_j$, and exchanged $m_{\text{ex}}$ in Eq.~\eqref{eq:cond}. Requiring the condition to hold in all cases leads to the constraint
\begin{equation}
\left(\sqrt 2-1\right)m_b<m_a<\frac{3+\sqrt{17}}{2}\,m_b\,.
\end{equation}
The lower bound originates from assigning the distinct mass to the exchanged line ($m_i=m_j=m_b$, $m_{\text{ex}}=m_a$), for which Eq.~\eqref{eq:cond} factorizes as $(x+1)(x^2+2x-1)>0$ with $x=m_a/m_b$, giving $x>\sqrt 2-1$. The upper bound originates from assigning it to the incoming spectator ($m_i=m_a$, $m_j=m_{\text{ex}}=m_b$), for which $R(s_{\text{thr}})=2m_a^2/[(m_a+m_b)(m_a+2m_b)]$ and $R(s_{\text{thr}})<1$ requires $m_a<(3+\sqrt{17})\,m_b/2$. The remaining assignment, with the distinct mass as the outgoing spectator, imposes no constraint.

For the $K\bar K\pi$ system considered in this work, one has $m_a=m_\pi$ and $m_b=m_K$, and the above condition is clearly not satisfied. Therefore, the integration contour discussed above cannot be safely applied in this case. In contrast, for the $\eta NN$ system studied in Ref.~\cite{Fix:2001cz}, the condition is fulfilled. As a result, although that system does not consist of three identical particles, the same integration contour can still be employed there without difficulty.\footnote{While the discussion above only establishes the condition at threshold, one can further verify that $R(s)<1$ remains satisfied throughout the energy region considered in Ref.~\cite{Fix:2001cz}.}

In summary, the applicability of this method remains rather restricted, and it is most naturally suited to systems with equal or nearly equal masses. The advantage of Method~2 is therefore its generality: it does not rely on the special mass-dependent condition $R(s)<1$ and never requires the potential to be evaluated on the second RS. Its drawback is the numerical cost. For each fixed $s$, $\text{SMC}_{i}(s,p^\prime)$ depends explicitly on $p^\prime$, and for each point $p$ on that contour one needs a separate AIC to evaluate $\mathcal A^J_i(s,p)$. This makes the method rigorous but expensive.

\subsection{Other singularities}

In addition to the three-body cuts, other singularities must also be respected when constructing the contours. The condition $E_{\boldsymbol{p}+\boldsymbol{p}^\prime,\text{ex}}=0$ in Eq.~\eqref{eq:OPE} leads to the $s$-independent singularities
\begin{equation}\label{eq:lhcex}
p_{\pm,\text{ex}}\left(p^\prime,z\right)=-p^\prime z\pm\sqrt{p^{\prime2}\left(z^2-1\right)-m_\text{ex}^2}\,.
\end{equation}
As in the three-body-cut case, the corresponding singular sets must not intersect the chosen contour. In practice, however, they usually lie far away from the relevant integration region and do not cause difficulties once the contour deformation is chosen properly.

Another source of singularities comes from the poles of the isobar propagator. Through the kinematic mapping of Eq.~\eqref{eq:kin}, poles in the $\sigma$ plane are mapped into isolated poles in the spectator momentum plane. Since these are isolated points rather than extended cuts, they are straightforward to avoid numerically.

Eventually, $\mathcal A_{j}^J(s,p^\prime)$ is rigorously defined for all real $p^\prime$. However, the AAA method is sufficient for the computation of physical observables such as the lineshape, so in the main text we do not adopt the contour deformation approach.

\bibliographystyle{JHEP}
\bibliography{refs}

\end{document}